\documentclass[aps,prd,article,preprint,onecolumn,superscriptaddress,tightenlines,nofootinbib,eqsecnum,preprintnumbers,longbibliography,12pt]{revtex4-1}
\pdfoutput=1
\usepackage{amsmath,latexsym,amssymb,graphicx,ifpdf,slashed,color,hyperref,url,cancel,comment}
\usepackage[T1]{fontenc}
\usepackage{relsize}
\usepackage{ulem}

\definecolor{niceblue}{rgb}{0.1, 0.3, 0.8} 
\definecolor{BrickRed}{rgb}{0.8, 0.25, 0.33}
\hypersetup{colorlinks,citecolor= BrickRed, linkcolor= niceblue, urlcolor=niceblue}

\begin{document}

\preprint{}

\title{Refractive neutrino masses, ultralight dark matter and cosmology}

\author{Manibrata Sen}
\email{manibrata@mpi-hd.mpg.de}
\affiliation{Max-Planck-Institut für Kernphysik, Saupfercheckweg 1, 69117 Heidelberg, Germany}
\author{Alexei Y. Smirnov}
\email{smirnov@mpi-hd.mpg.de}
\affiliation{Max-Planck-Institut für Kernphysik, Saupfercheckweg 1, 69117 Heidelberg, Germany}
\affiliation{School of Physics, Korea Institute for Advanced Study, Seoul, 02455, Republic of Korea}


\begin{abstract}

We consider in detail a possibility that the observed neutrino oscillations 
are due to refraction on ultralight scalar boson dark matter.
We introduce the refractive mass squared, $\tilde{m}^2$,  and study its properties: dependence on neutrino energy,
state of the background, etc. If the background is in a state of cold gas of particles, $\tilde{m}^2$
shows a resonance dependence on energy. Above the resonance ($E \gg E_R $), we find that $\tilde{m}^2$ 
has the same properties as usual vacuum mass squared.
Below the resonance, $\tilde{m}^2$ decreases with energy, which (if realised) allows us to avoid the cosmological 
bound on the sum of neutrino masses. Also, $\tilde{m}^2$ may depend on time. 
We consider the validity of the results: effects of multiple interactions with
scalars, and modification of the dispersion relation. We show that for values of parameters of the system required to
reproduce the observed neutrino masses, perturbativity is broken at low energies, which border above the resonance.
If the background is in the state of coherent classical field, the refractive mass does not depend on energy explicitly but may show
time dependence. It coincides with the refractive mass in a cold gas at high energies. 
Refractive nature of neutrino mass can be tested by searches of its dependence on energy and time.
\end{abstract}

\maketitle

%%%%%%%%%%%%%%%%%%%%%%%%%%%%%%%%%%%%%%%%%
\section{Introduction}
%%%%%%%%%%%%%%%%%%%%%%%%%%%%%%%%%%%%%%%%%
%%%%%%%%%%%%
One of the greatest achievements in particle physics is ``... the discovery of neutrino oscillations,
which shows that neutrinos have mass''. 
Indeed, oscillations of atmospheric neutrinos~\cite{Kajita:2016cak} and adiabatic 
conversion of solar neutrinos~\cite{McDonald:2016ixn} have been discovered.
But how do we know that the neutrino mass is behind the oscillations and conversion?
Smallness of the observed neutrino mass and large mixing may testify that its nature  
differs from nature of other fermions.

In 1978, Wolfenstein proposed the oscillations of massless neutrinos~\cite{Wolfenstein:1977ue}. 
For this, he introduced point-like four-fermion interactions (what we call now NSI) 
which generate effective potentials experienced
by neutrinos. 
The potentials mix neutrinos and give splitting of energy eigenvalues of propagation. 
Presumably, the interactions are due to exchange of heavy mediators.  
Therefore the  potentials, and consequently, oscillation effects do not depend on neutrino energy.

However, the energy dependence of oscillations was found in experiments! Furthermore, the dependence 
is in agreement with the presence of the mass 
term in the Hamiltonian of evolution: 
$$
H \sim \sqrt{p^2 + |m|^2} \approx p + \frac{|m|^2}{2E}\,,
$$
(in 3 $\nu$ case $m^2 \rightarrow M M^{\dagger}$,  where $M$ is $3\times 3$ mass matrix). 
It is this energy dependence of the oscillation effects 
which provides a convincing argument that the neutrino mass is behind oscillations.

Still, this is not the end of the story.
Notice that oscillations of relativistic neutrinos probe mass-squared and not the mass. Furthermore, 

(i) mass changes the chirality of fermions, while the mass-squared does not.

(ii) The mass operator of neutrinos has a gauge charge and appears as a result of symmetry breaking. 
In contrast, modulus of mass squared is gauge invariant and does not require symmetry breaking. 
Therefore, in oscillations there is no direct probe of mass,
and any contribution to the Hamiltonian of evolution which has an $A/E$ form
with a constant $A$ can reproduce the oscillation data.

In fact, at high enough energies the potential generically has a $1/E$ dependence.  
In the Standard Model (SM), the potentials have the $1/E$ dependence above the threshold 
of production of the $W-$ and $Z-$ boson mediators, i.e., for $s \gtrsim m_{\rm mediator}^2$~\cite{Lunardini:2000swa}. 
For neutrino scattering on electrons due to $W-$boson exchange,
this requires the neutrino energy in the laboratory frame
$E \gtrsim m_W^2/2m_e = 6 \cdot 10^6$ GeV. 
If mediator is light, the $1/E$ dependence of the potential shows up at low observable energies.
This opens up a possibility to substitute, to some extent, the vacuum mass term in the Hamiltonian
by the energy-dependent potential~\cite{Choi:2019zxy, Ge:2020ffj, Ge:2020xkm, Choi:2020ydp,Chun:2021ief}. 

Since oscillations are observed in ``vacuum'',
where interactions with matter can be neglected, the scatterers should be new light particles
that fill the space, thus being a component of the Dark Matter (DM).
For example, fuzzy DM fits all these requirements~\cite{Hu:2000ke,Hui:2016ltb}.
In this connection, refraction effects of neutrinos on very light target particles due to exchange
of light mediators were studied and the effective potentials were computed
\cite{Choi:2019zxy,Choi:2020ydp,Smirnov:2021zgn}. 
The phenomenology of neutrino-ultralight DM interactions 
has been explored extensively~\cite{Berlin:2016woy,Krnjaic:2017zlz,Brdar:2017kbt,Capozzi:2018bps,Dev:2020kgz,
Losada:2021bxx,Huang:2021kam,Chun:2021ief,Dev:2022bae,Huang:2022wmz,Davoudiasl:2023uiq,Losada:2023zap,Gherghetta:2023myo}.

In this paper we address the following questions: 
can the potential with $1/E$ dependence substitute the mass completely?
Can one distinguish the usual mass and potential in oscillations or in some other way ?

One fundamental difference exists between neutrino mass generation by vacuum expectation value (VEV) and that by refraction.
Refractive mass is proportional to the number density of DM particles: 
$m^2 \propto n$, and therefore the neutrino mass would increase in the past 
with redshift as $(z+1)^{3/2}$. If the present ($z =0 $) neutrino mass  
bound is considered for the heaviest active neutrino, then in the epoch 
of photon decoupling $(z \sim 1000)$, the neutrino mass would become $O(10)$ eV (see details in the text). 
This violates the cosmological bound on the sum of neutrino masses 
from observations of the cosmic microwave background (CMB), 
as well as bounds from structure formation~\cite{Planck:2018vyg}. 
The cosmological bound still allows contribution to the neutrino mass at the level $3\cdot  10^{-3}\,{\rm eV}$, 
affecting solar neutrinos~\cite{Berlin:2016woy}.

In this paper, we consider a realistic scenario of the refractive mass  
with three active neutrinos which can reproduce all the oscillation data. 
We further elaborate on the properties of refractive neutrino masses. 
It will be shown that by 
appropriate selection of the resonance energy, the cosmological bound may be satisfied. 
Dependence of the refractive masses on he state of medium is discussed. We also study perturbativity 
violation due to multiple interactions of neutrino with the dark matter particles.

The paper is organized as follows. In sect.\,\ref{sec:RefMassPart}, we introduce the refractive neutrino 
mass in a cold bath of scalar particles and study its properties. 
We show that with appropriate choice of parameters the existing experimental data on neutrino oscillations can be 
explained. In sect.\,\ref{sec:validity}, the conditions of perturbativity violation and validity of the results are considered.
In sect.\,\ref{sec:RefMassField}, we discuss the refraction of neutrinos in a classical field that describes 
a coherent state of scalar particles, and the corresponding contribution to the neutrino mass matrix. 
We discuss constraints arising from astrophysical sources, terrestrial laboratories 
as well as the early Universe in sect.\,\ref{sec:AstroLab} and determine the viable parameter space. 
Finally, we summarize our results and  conclude in sect.\,\ref{sec:Concl}.
%%%%%%%%%%%%%%%%%
%%%%%%%%%%%%%%%%%

\section{Refractive masses of neutrinos in cold particle bath. }
\label{sec:RefMassPart}
%%%%%%%%%%%%%%%%%%%%%%%%%%%%%%%%%%%%%%%%%%%%%%%%%%%%%%%%%%%%%%%%%%%%%%%%%%
%%%%%%%%%%%
%%%%%%%%%%%
\subsection{Effective potential and refractive mass}
%%%%%%%%%%%%%%%%%%%%%%%%%%%%%%%%%%%%%%%%%%%%%%%%%%%%%%%%%%%%%%%%%%%%%%%
Let us consider the propagation of massless neutrinos $(\nu_e, \nu_\mu, \nu_\tau)$ 
in a medium composed of ultralight scalar bosons 
$\phi$ and their antiparticles $\bar{\phi}$ with number densities $n_\phi$ and $\bar{n}_\phi$
correspondingly. 
These ultralight scalars can act as the cold DM, or compose a part of the DM. 
Neutrinos scatter on the scalars via exchange of light fermionic mediators $\chi_k$ due to the Yukawa 
couplings,
\begin{equation}
    \mathcal{L}\supset \sum_{\alpha = e, \mu, \tau} \sum_k g_{\alpha k} \bar{\chi}_{ k R}\nu_{\alpha L} \phi^* +  
m_{\chi k} \bar{\chi}_{ k R} \chi_{ k L}  + {\rm h.c}  \,.
    \label{eq:Lag}
\end{equation}
There are two possible cases: Dirac fermion mediators and Majorana mediators: in the latter case 
$\chi_{ k L} \rightarrow (\chi_{ k R})^c$.   
As we will see, in order to explain the oscillation data, 
at least two mediators are needed with different couplings $g_{\alpha k}$.

\begin{figure}[!t]
\includegraphics[width=0.9\textwidth]{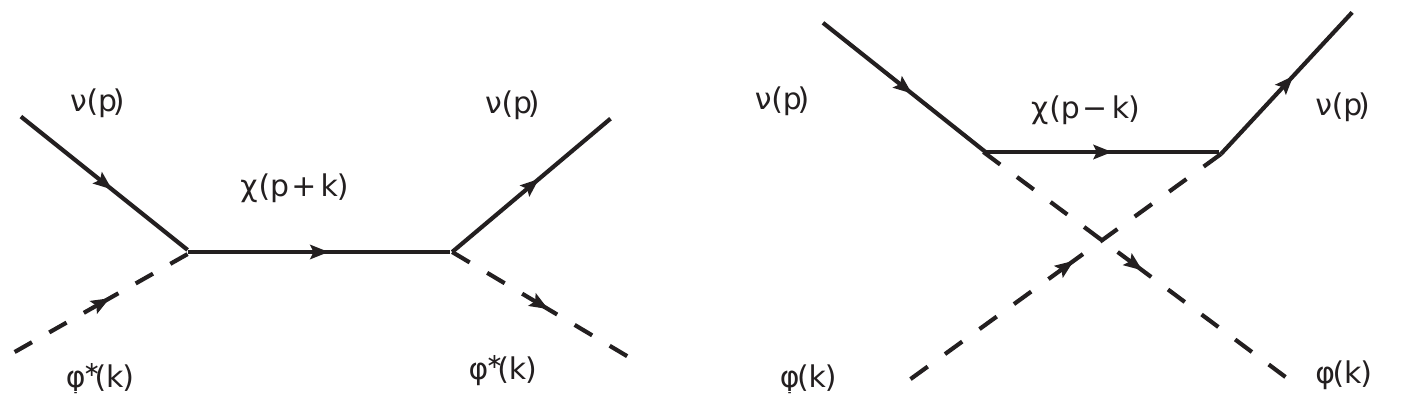}
\caption{ Feynman diagrams for scattering of neutrinos on the $\phi$ background. Left: $s-$channel. Right: $u-$channel.}
\label{fig:potential}
\end{figure}

The elastic forward scattering $\nu \phi \rightarrow \nu \phi$ 
in the cold gas of $\phi-$particles produces the effective potential through the $s-$channel 
and $u-$channel scattering as shown in Fig.\,\ref{fig:potential}~ 
\cite{Choi:2019zxy,Choi:2020ydp,Smirnov:2021zgn}, 
\begin{equation}
    V_{\alpha\beta} = \sum_k g_{\alpha k} g_{\beta k}^*\,\left[
\frac{\bar{n}_\phi(2 E m_\phi - m_{\chi k} ^2)}{(2 E  m_\phi - m_{\chi k}^2)^2 + (m_\chi \Gamma_{\chi k})^2} +  
\frac{n_\phi}{2 E  m_\phi + m_{\chi k}^2}  
\right]\,,
    \label{eq:PotPhi}
\end{equation}
where $E$ is the neutrino energy,  
$m_{\phi}$  is  the mass of $\phi$, 
and $\Gamma_{\chi k} \equiv \sum_\alpha g_{\alpha k}^2 m_{\chi k}/(8\pi)$ 
is the total decay rate of the mediator $\chi_k$. 
We consider non-relativistic $\phi$ and
neglect its mass-squared in the denominator in (\ref{eq:PotPhi}).

The first term in (\ref{eq:PotPhi}) corresponds to $\chi_k$ exchange in the $s-$channel and for 
$m_{\chi k} > m_{\phi}$ it has a resonance character. For simplicity, we will introduce two mediators with 
nearly the same masses: $m_{\chi k} \approx m_\chi$. 
According to (\ref{eq:PotPhi}), for $\phi$ at rest,  the resonance energy equals
\begin{equation}
E_R\simeq \frac{m_\chi^2}{2 m_\phi}. 
\label{eq:resen}
\end{equation}
The second term in (\ref{eq:PotPhi}) is due to $\chi_k$ exchange in the $u-$channel. 
For antineutrinos, the potential can be obtained from (\ref{eq:PotPhi}) by interchanging 
$n_\phi \leftrightarrow \bar{n}_\phi$.

In the $\phi-$background,  the Hamiltonian of evolution of massless neutrinos  
$(\nu_e, \nu_\mu, \nu_\tau)$  is given by  
\begin{equation}
H \simeq 
%\frac{M^2}{2E} + 
{V}, 
\end{equation}
where 
${V}  = || V_{\alpha\beta} ||$  
is the matrix of potentials. We introduce the refractive neutrino mass squared 
as 
$$
\tilde{m}_{\alpha\beta}^2 \equiv 2 E V_{\alpha \beta}\,,
$$  
or, in a matrix form as
\begin{equation}
\tilde{m}^2 = 2EV. 
\label{eq:effmass}
\end{equation}
Then the Hamiltonian can be written  in the form of the vacuum term:
\begin{equation}
H \simeq \frac{\tilde{m}^2}{2E}\,.
\end{equation}

Let us study the properties of the refractive masses-squared 
$\tilde{m}_{\alpha\beta}^2$.
In terms of the dimensionless parameter $y \equiv E /E_R$, the refractive mass (\ref{eq:effmass}), 
with $V$ given in (\ref{eq:PotPhi}), can be written as 
\begin{equation}
 \tilde{m}_{\alpha\beta}^2 = 2 y E_R \sum_k V_{\alpha \beta k}^0\,\left[
\frac{(1-\epsilon)(y-1)}{(y-1)^2+\tilde{\xi}_k^2} +  
\frac{1+\epsilon}{1+y} 
\right]\, .
    \label{eq:RedPotPhi}
\end{equation}
Here 
\begin{equation}
 V_{\alpha \beta k}^0 \equiv  \frac{g_{\alpha k} g_{\beta k}^*}{2m_\chi^2}(n_\phi+\bar{n}_\phi),    
\label{eq:vzero}
\end{equation}
$\epsilon$ is the C-asymmetry of background: 
\begin{equation}
\epsilon \equiv \frac{n_\phi-\bar{n}_\phi}{n_\phi+\bar{n}_\phi}\,,~~~  (\epsilon = -1 \div 1), 
\label{eq:asym}
\end{equation} 
and  $\tilde{\xi}_k \equiv \Gamma_{\chi k}/m_\chi$.  
Since the couplings are very small, we have  $\tilde{\xi}_k \ll 1$,  
and therefore $\tilde{\xi}_k$ can be neglected everywhere apart from a very narrow region around the resonance,  
$y = 1$. The energy dependent factor in the brackets of (\ref{eq:RedPotPhi})  
does not depend on mediator type $k$ and becomes universal for all the contributions.  
Then using explicit expressions for $V_{\alpha \beta k}^0$ and $E_R$, 
the matrix of refractive mass-squared can be written as   
\begin{equation}
\tilde{m}^2 = \lambda \frac{n_\phi + \bar{n}_\phi}{m_\phi}~ \frac{y (y - \epsilon)}{y^2 - 1}\,,
\label{eq:potenwxi}
\end{equation}
where $\lambda$ is the matrix of couplings:
$$
\lambda = ||\lambda_{\alpha \beta}||,  ~~~~~  \lambda_{\alpha \beta} = \sum_k  g_{\alpha k} g_{\beta k}^*\,.
$$
Introducing vectors of Yukawa couplings $\vec{g}^T_k \equiv (g_{ek}, g_{\mu k}, g_{\tau k})$, $\lambda$ can be expressed as
\begin{equation}
\lambda =  \sum_k \vec{g}^T_k \times \vec{g}_k\,.
\end{equation}
Notice that $\tilde{m}^2$ in (\ref{eq:potenwxi})  does not depend on the mediator mass explicitly. 
Defining 
\begin{equation}
\tilde{m}^2_{\rm asy}\equiv \lambda \frac{n_\phi + \bar{n}_\phi}{m_\phi}, 
\label{eq:minf}
\end{equation}
the effective mass-squared (\ref{eq:potenwxi}) becomes
\begin{equation}
\tilde{m}^2 = \tilde{m}^2_{\rm asy} \frac{y (y - \epsilon)}{y^2 - 1}\,.
\label{eq:potenwxi1}
\end{equation}
Since the contribution of $\phi$ to the energy density of the Universe
\begin{equation}
\rho_\phi = m_\phi (n_\phi + \bar{n}_\phi )\,,
\end{equation}
we can rewrite the refractive masses (\ref{eq:minf})  as 
\begin{equation}
\tilde{m}^2_{\rm asy} =  \lambda \frac{\rho_\phi}{m_\phi^2}\, .      
 \label{eq:mfromrho}
\end{equation}

%%%%%ffff2%%%%%%%%%%%%%%%%%%%%%%%%%
\begin{figure}[!t]
\includegraphics[width=0.7\textwidth]{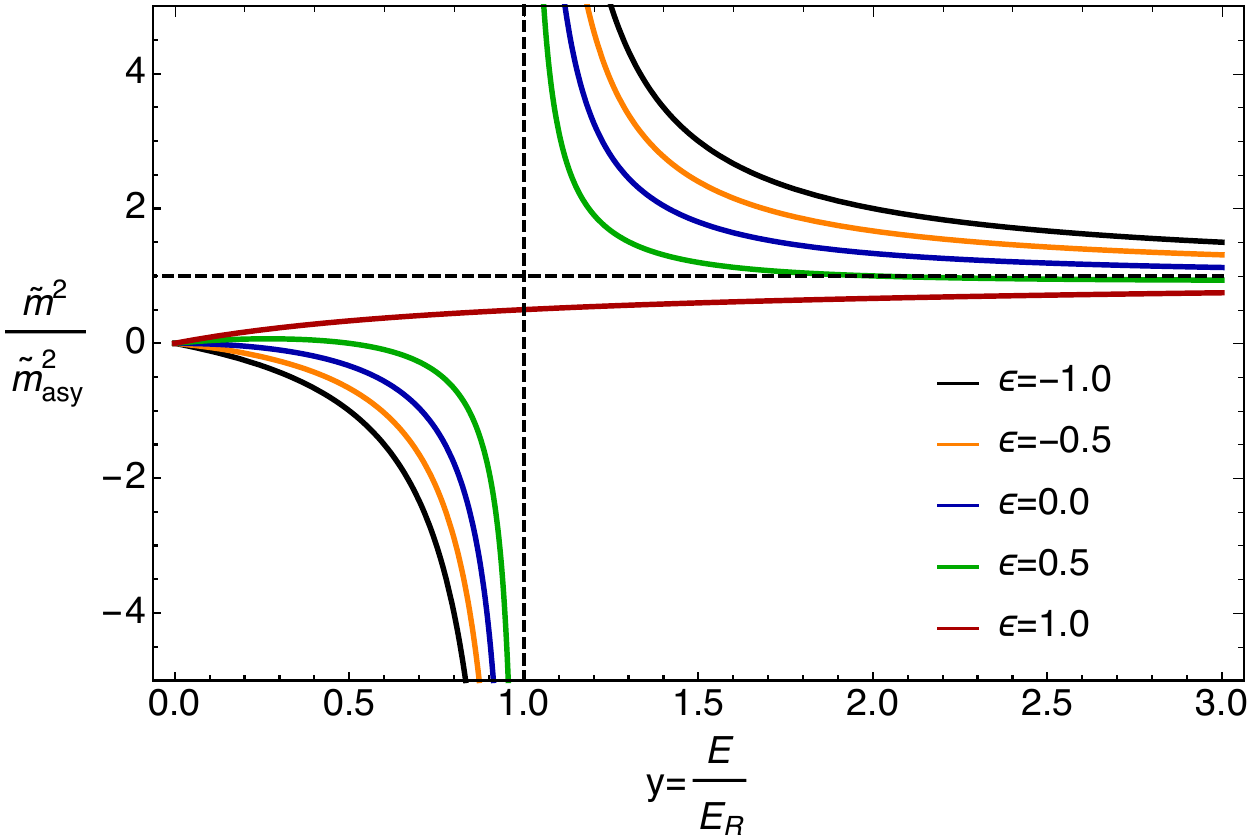}~~~
\caption{The rescaled refractive mass squared  $\tilde{m}^2/\tilde{m}^2_{\rm asy}$
as function of rescaled energy, $y$, for different values of the asymmetry $\epsilon$.}
\label{fig:Delmeff_g_mphi}
\end{figure}
\begin{figure}[!t]
\includegraphics[width=0.7\textwidth]{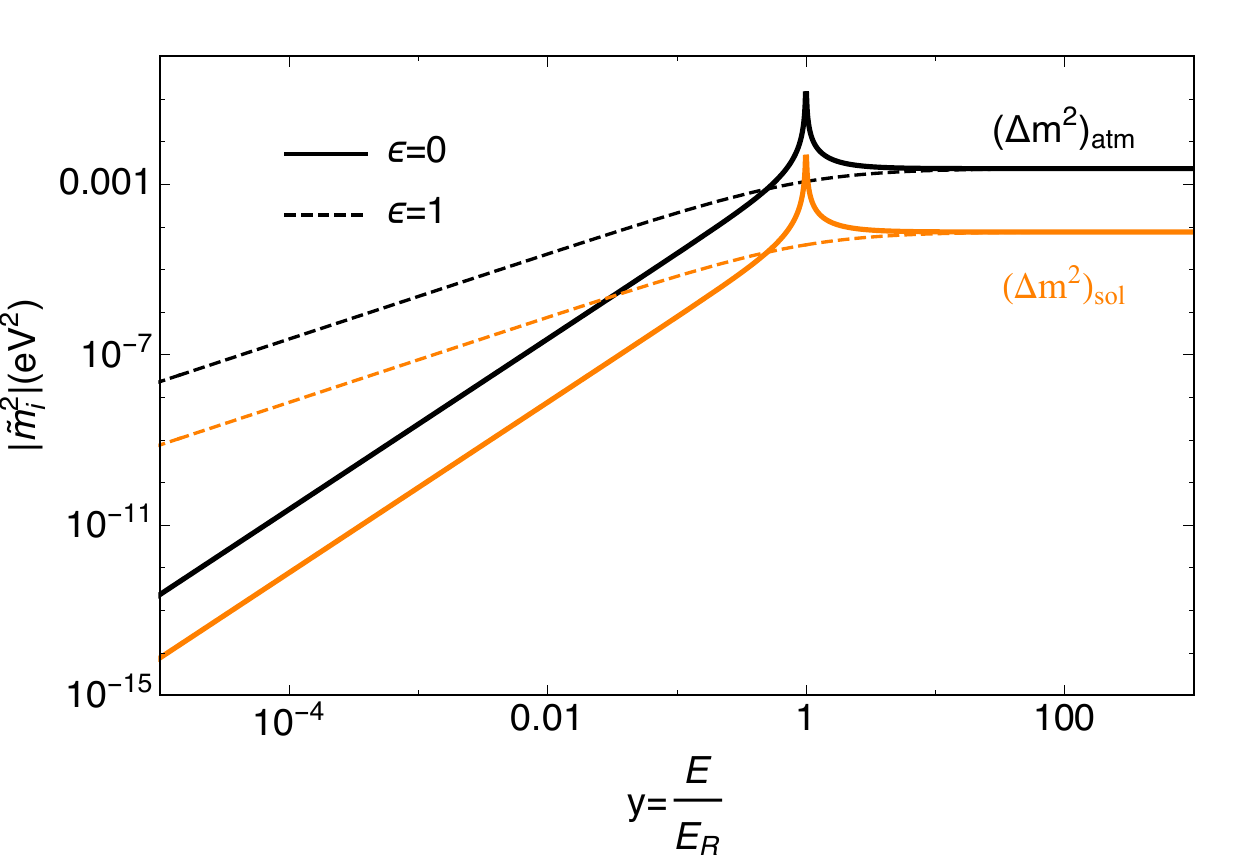}
\caption{The moduli of eigenvalues of the refractive mass-squared matrix, $|\tilde{m}^2|$, 
as a function of rescaled energy. 
In asymptotics, $y \gg 1$, 
they can reproduce the solar and the atmospheric neutrino mass-squared differences.  }
\label{fig:Delmeff_g_mphi2}
\end{figure}

 
%%%%%%%%%%%%%%%%%%%%%%%%%%%%%%%%%%%%%%%%%%%%%%%%%%%%%
The dependence of (\ref{eq:potenwxi1}) is shown in 
Fig.\,\ref{fig:Delmeff_g_mphi} and  Fig.\,\ref{fig:Delmeff_g_mphi2}. In
Fig.\,\ref{fig:Delmeff_g_mphi}, we present the ratio $\tilde{m}^2/\tilde{m}^2_{\rm asy}$  as a function 
of the rescaled energy $y$ for different values of the asymmetry  $\epsilon$. Fig.\,\ref{fig:Delmeff_g_mphi2} shows the 
dependence of the absolute values of the refractive mass squared 
$|\tilde{m}^2|$ on $y$.   
The  mass squared has the following properties:

\begin{enumerate}

    \item For very low energies,  $y \ll 1$,  we have  
\begin{equation}
\tilde{m}^2/ \tilde{m}^2_{\rm asy} = y(\epsilon - y).
\label{eq:massvar}
\end{equation}
When $y \rightarrow 0$, it decreases as $y\epsilon$. For $\epsilon = 1$, the potential 
reproduces the Wolfenstein potential $V = 2 V_0 = (g^2/m_\chi^2) n_\phi$. 
In the C-symmetric medium,  $\epsilon  = 0$, the mass squared  has an additional suppression factor $y$: 
$$
\tilde{m}^2/ \tilde{m}^2_{\rm asy}= -  y^2 . 
$$

\item At resonance, $y=1$,  
the contribution from the $s-$channel diagram is zero, and it changes sign as the energy 
falls below the resonance energy.  The contribution from the $u-$channel slightly 
shifts the pole to $\tilde{m}^2= \tilde{m}^2_{\rm asy}(1+\epsilon)/2$.
   
\item In the limit of high energies, $y \gg (1, \epsilon)$, 
with increase of $y$, the mass converges to the asymptotic value: 
$\tilde{m}^2 \rightarrow \tilde{m}^2_{\rm asy}$.
According to (\ref{eq:potenwxi1}), 
the masses reduce to 
\begin{equation}
\tilde{m}^2/ \tilde{m}^2_{\rm asy}=  
\left\{
\begin{matrix}
1 - \frac{\epsilon}{y}~,    ~~~~~\epsilon \neq 0\,. \\
1 + y^{-2}~, ~~~ \epsilon = 0\,.
\end{matrix}
\right.
\label{eq:potenhe}
\end{equation}
The convergence is faster for zero asymmetry.

\item For antineutrinos, the refractive mass squared can be obtained by changing 
the sign of C-asymmetry: $\epsilon \rightarrow - \epsilon$. While at low energies, 
this changes the sign of $\tilde{m}_{\alpha \beta}^2$, at high energies, $y \gg 1$,  
the mass $\tilde{m}_{\alpha \beta}^2$ is nearly the same for neutrinos and antineutrinos. 
This contradicts the statement in \cite{Ge:2019tdi}.

\end{enumerate}

%%%%%%%%%%%%%%%%%ffff3 %%%%%%%%%%%%%
%%%%%%%%%%%%%%%%%%%%%%%%%%%%%%%%%%%%%%%%%%%%%%%%%%%%%
Thus,  at high energies, $E \gg E_R$,  the refractive mass-squared  $\tilde{m}^2$ has  the same 
properties as the usual vacuum mass-squared $m^2$: no dependence on energy, the same for 
neutrinos and antineutrinos.  
This is not surprising: essentially,  in this case, 
the VEV (classical scalar field at the lowest energy state) is substituted by scalar particles 
-- quanta of the scalar field. Furthermore, at high enough density  
(occupation number), the sea of scalar particles can be 
treated as a classical scalar field, as we will discuss later.   

Similarly, due to interaction with $\phi$ and $\nu$, the fermion $\chi$ acquires a refraction 
mass $\tilde{m}_{\chi k}^2$  which is of the same order as the neutrino refraction mass.

%%%%%%%%%%%%%%%%%%%%%%%%%%%%%%%%%%%%%%%%%%%%%%%%%%%%%%%%%%%%%%%%%%%%%
\subsection{Refractive mass and the neutrino oscillation data}
%%%%%%%%%%%%%%%%%%%%%%%%%%%%%%%%%%%%%%%%%%%%%%%%%%%%%%%%%%%%%%%%%%
\label{sec:refmass}
%%%%%%%%%%%%%

For a single mediator, the matrix of couplings   
$
\lambda =  \vec{g}^T \times \vec{g}
$
has rank 1, that is, in the $3\nu-$case, only one neutrino gets an effective 
mass and only two mixings are defined.  Therefore, the second mediator should 
be introduced with different set of couplings to active neutrinos.  
In this case, 6 real couplings can explain 6 real parameters: 
three mass squared differences and three mixing angles. 
For complex coupling constants, the CP-violation phase can be obtained. 
Moreover, 5 couplings are enough, so that  
one of the couplings  can be set to zero to reproduce two $\Delta m^2$  and three angles.   

Let us fix values of the parameters $g_{\alpha k}$, $m_\phi$  and $m_\chi$ 
to fit the neutrino oscillation data with refraction mass squared (\ref{eq:RedPotPhi}).  
For this, we consider the matrix of active neutrinos
$\tilde{m}^2_{\alpha \beta}$ of the nearly tribimaximal (TBM) form~\cite{Harrison:2002er}. 
For
\begin{equation}
g_{e 1} = g_{\mu 1} = g_{\tau 1} = g_1, ~~ \, \, \,
g_{e 2} = 0, \, \, ~~  g_{\mu 2} = - g_{\tau 2} = g_2\,,
\label{eq:gall}
\end{equation}
the elements of the $3\times3$ matrix $\tilde{m}^2_{\alpha \beta}$ become
\begin{equation}
\tilde{m}^2_{e \alpha} = g_1^2\,\frac{\rho_\phi}{m_\phi^2}\,, ~~~~ \alpha = e,  \mu, \tau, 
\label{eq:ealpha}
\end{equation}
\begin{equation}
\tilde{m}^2_{\mu \mu} = \tilde{m}^2_{\tau \tau} =
(g_1^2 + g_2^2)\frac{\rho_\phi}{m_\phi^2}, ~~~~
\tilde{m}^2_{\mu \tau} = \tilde{m}^2_{\tau \mu} = (g_1^2 - g_2^2)\frac{\rho_\phi}{m_\phi^2}\,.
\label{eq:mutau}
\end{equation}
This exactly reproduces the TBM structure. Since the lightest neutrino mass is zero, 
the values of the parameters can be connected to the observed mass-squared differences as
\begin{equation}
g_1^2 \frac{\rho_\phi}{m_\phi^2}= \frac{1}{3} \Delta m_{\rm sol}^2\,, ~~~~
g_2^2 \frac{\rho_\phi}{m_\phi^2} =  \frac{1}{2} \Delta m_{\rm atm}^2 .
\label{eq:conn}
\end{equation}
For normal mass ordering, we have from (\ref{eq:conn}),
\begin{eqnarray}
g_{1} &=& m_\phi \sqrt{\frac{\Delta m_{\rm sol}^2}{3\rho_\phi}} = 
3.2\cdot 10^{-10} \left(\frac{m_\phi}{10^{-10}\,{\rm eV}}\right)\,\left(\frac{\Delta m_{\rm sol}^2}{7.5\cdot 10^{-5}\,{\rm eV}^2}\right)^{1/2}\,\left(\frac{\rho_\odot}{\rho_\phi}\right)^{1/2} \,,
\nonumber\\
g_{2} &=& m_\phi \sqrt{\frac{\Delta m_{\rm atm}^2}{2\rho_\phi}}= 
2.2\cdot 10^{-9} \left(\frac{m_\phi}{10^{-10}\,{\rm eV}}\right)\,\left(\frac{\Delta m_{\rm atm}^2}{2.3\cdot 10^{-3}\,{\rm eV}^2}\right)^{1/2}\,\left(\frac{\rho_\odot}{\rho_\phi}\right)^{1/2} \,.
\label{eq:Delmsq}
\end{eqnarray}
For numerical estimation, we assumed that $\phi$ compose the entire local DM energy density $\rho_\odot$, therefore, $\rho_\phi = \rho_\odot = 0.3\, {\rm GeV cm}^{-3}$. 
The dependences (\ref{eq:Delmsq})  are shown in Fig. \ref{fig:bounds}
together with various bounds, which we will discuss later.
The relations (\ref{eq:Delmsq}) 
are based on asymptotic values of the effective masses and, therefore, practically do not depend
on $m_\chi$. 

The bounds on neutrino-scalar interactions coupling $g$ follow from laboratory experiments.   
Invisible decay of the Higgs~\cite{ATLAS:2016neq}, invisible Z-decay~\cite{Electroweak:2003ram}, 
meson decays~\cite{Pasquini:2015fjv}, and tau decays~\cite{Brdar:2020nbj} 
give bounds  which are weaker than $g \sim 10^{-5}$. 
They are superseded by astrophysical and cosmological bounds.
The strongest bound follows from cooling of stars: 
$g_i < 10^{-7}$~\cite{Farzan:2018gtr,Dev:2020eam}. Therefore according to (\ref{eq:Delmsq}),
we obtain an upper bound on the mass of scalar
\begin{equation}
m_\phi < 5 \cdot 10^{-9}\, {\rm eV} \left(\frac{g_2}{10^{-7}} \right)\,.
\label{eq:mphibound}
\end{equation}

The upper bound on the resonance energy follows from
non-observation of energy dependence of oscillation parameters, 
namely, increase of the effective mass squared
with decrease of energy (see Fig. \ref{fig:Delmeff_g_mphi}). 
The lowest energies of detected neutrinos are about 0.2 MeV (the solar pp-neutrinos).
At these energies, oscillations are averaged and therefore are insensitive to $\Delta m^2$. 
The sensitivity appears in reactor neutrinos with $E \sim$ MeV. 
The accuracy of determination of $\Delta m^2$ is about $10\%$
and data are in agreement with constant $\Delta m^2$.
Also no dependence on energy has been found in a wide energy range
(from IC-Deep Core down to reactor neutrinos) and all the extractions of $\Delta m_{\rm atm}^2$ 
are in agreement within $10\%$.
Therefore, according to (\ref{eq:potenhe}), $y^2 > 10$ for
$\epsilon = 0$, and $y > 10$ for $|\epsilon| = 1$ are required.
Taking the lowest testable energy 1 MeV, we find the upper bound on resonance energy $E_R < 0.1$ MeV.

%%%%%%%%%%%%%%%%%%%%%%%%%%%%%%%%%%%%%%%%%%%%%%%%%%%%%%%
\begin{figure}[t]
\includegraphics[width=0.7\textwidth]{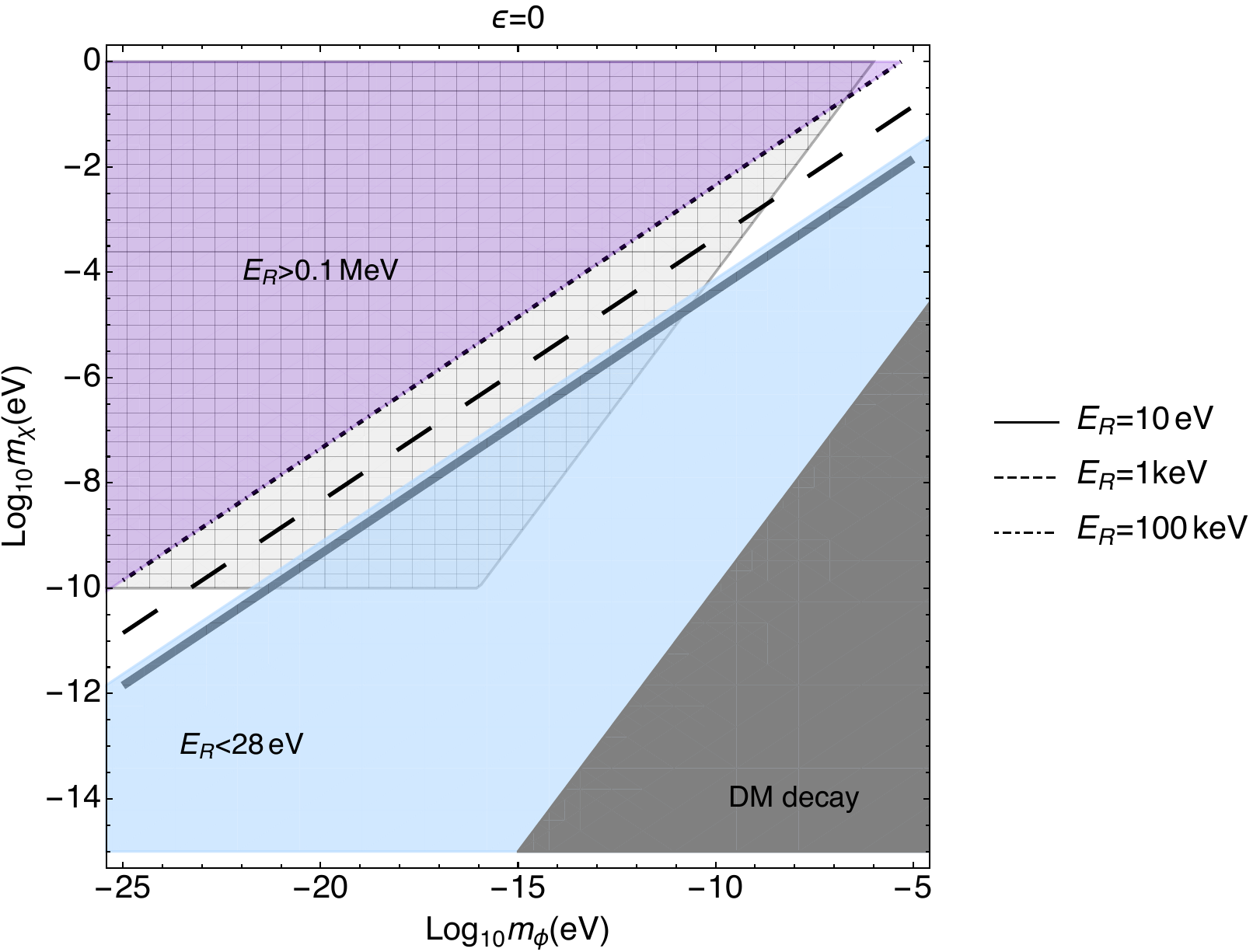}
\caption{The allowed region in the $m_\chi-m_\phi$ plane. 
The pink shaded region corresponds to $E_R>0.1\,{\rm MeV}$ and is excluded by oscillation data. In the blue shaded region, 
$E_R<28\,{\rm eV}$ and this is excluded by cosmological considerations (see discussion in sect.\,\ref{sec:cosmomass}). The gray shaded 
region corresponds to $m_\phi>m_\chi$, thereby making the DM unstable. The hatched region is excluded by $\nu-\chi$ oscillations of solar neutrinos (see sect.\,\ref{sec:nuchiosc}). }
\label{fig:Eres_mchi_mphi}
\end{figure} 
%%%%%%%%%%%%%%%%%%%%%%%%%%%%%%%%%%%%%%%%%%%%%%%%%%%%%%%%%%%%%%
%

The mass of mediator  is determined by $m_\phi$ and resonance energy  (\ref{eq:resen}) through
\begin{equation}
m_\chi =  \sqrt{2 m_\phi E_R } \,.
\label{eq:mchi}
\end{equation}
From this equation and (\ref{eq:mphibound}), we find 
\begin{equation}
m_\chi < 0.03\, {\rm eV}\,
\left(\frac{m_\phi}{5 \cdot 10^{-9}\, {\rm eV}} \right)^{1/2} 
\left(\frac{E_R}{0.1\, {\rm MeV}  } \right)^{1/2}\,.
\label{eq:chimass}
\end{equation}
Fig.\,\ref{fig:Eres_mchi_mphi} shows $m_\chi$ as function of $m_\phi$ 
for different values of $E_R$. 

%\MS{The constraint arising from the upper(lower) bound on $E_R$ in pink (blue). We assume that the mixing between the neutrino and $\chi$ in this case is negligible, hence bounds arising from $\chi$ production from neutrinos are absent. These bounds will be discussed in the case of refraction on coherent background state described by classical field and is discussed in Sec.\,\ref{sec:RefMassField}.}

%%%%%%%%%%%%%%%%%%%%%%%%%%%%%%%%%%%%%5
\section{Scales and validity of results}
\label{sec:validity}
%%%%%%%%%%%%%%%%%%%%%%%%%%%%%%%%%%%%%%%%%%%%%%%%%

For small enough  $g/m_\phi$, and consequently small $\Delta m^2$, the results described above clearly work. However, with increase of $g/m_\phi$, various problems arise and the question we address in this section is whether we can still reproduce the observed values 
of the masses and avoid these problems.

\subsection{Scales in the problem}
%%%%%%%%%%%%%%%%%%%%%%%%%%%%%%%%%%%%%%%%%%%%%%%%%%%%%%%%%%%%%%%%%%%

Values of  $m_\phi$, $m_\chi$ and minimal $E_R$
estimated in sect.\,\ref{sec:refmass} determine several spatial 
scales, and consequently, the physical picture of the effects.
The Compton length of the scalar has a macroscopic size:
$$
\lambda_c = \frac{2\pi}{ m_\phi} > 2.5\cdot 10^5\,{\rm cm} \left(\frac{5\cdot 10^{-10}\,{\rm eV}}{m_\phi}\right)\,.
$$
The number density of scalars equals 
$$
n_\phi = \frac{\rho_\phi}{m_\phi} > 6 \cdot 10^{17}\,{\rm cm}^{-3}\left(\frac{5\cdot 10^{-10}\,{\rm eV}}{m_\phi}\right)\,,
$$
so that the distance between them 
$$
d \simeq n^{-1/3} \leq 1.2 \cdot 10^{-6}\,{\rm cm}\, \left(\frac{m_\phi}{5\cdot 10^{-10}\,{\rm eV}}\right)^{1/3}\,.
$$
In the Galaxy, the virialized velocity of DM is $v_{\rm vir} \sim 10^{-3} $, 
consequently, the de Broglie wavelength equals
$$
\lambda_{dB} = \frac{2\pi }{m_\phi v_{\rm vir}} > 2.5 \cdot 10^8\,{\rm cm}\,\left(\frac{5\cdot 10^{-10}\,{\rm eV}}{m_\phi}\right).
$$
Thus, the de Broglie length is bigger than typical baseline of laboratory experiments which means that
production and detection of neutrinos occurs within a single de Broglie wavelength of the $\phi$.
The radius of interaction is given by $m_\chi$:
$$
r_\chi = \frac{1}{m_\chi} =  6 \cdot 10^{-4}~ {\rm cm} \left(\frac{0.03~ {\rm eV}}{ m_\chi} \right)
$$ 
below resonance and 
$$
r_\chi \simeq (2E m_\phi)^{-1/2}
$$
above resonance. Thus, we have
\begin{equation}
\lambda_{dB} \gg r_\chi \gg d \gg \lambda_\nu,
\label{eq:scales4}
\end{equation}
where $\lambda_\nu$  is the wavelength of laboratory neutrinos.
In the following sections, we will discuss the consequences of the hierarchy 
in scales and how this can change the results of the computations (\ref{eq:PotPhi}),
which correspond to the standard $S$-matrix scattering formalism.
\subsection{Wavefunction renormalization and Perturbativity}
%%%%%%%%%%%%%%%%%%%%%%%%%%%%%%%%%%%%%%%%%%%%%%%%%%%%%%%%%%%%%%%%%
In the computation of (\ref{eq:PotPhi}), the propagator of the mediator $\chi$ equals   
\begin{equation}
\mathrm{i}\frac{\slashed{p}_\chi + m_\chi}{p_\chi^2 - m_\chi^2} = 
\mathrm{i}\frac{\slashed{p} \pm \slashed{k} + m_\chi}{p_\chi^2 - m_\chi^2}\,,
\label{eq:propagator}
\end{equation}
where the signs $+$ and $-$ are for the $s-$ and $u-$
channels correspondingly. 
The potential (\ref{eq:PotPhi}) is produced by the term $\slashed{k}$, 
which gives $m_\phi\gamma_0$ in the non-relativistic case. 
The contribution  $\slashed{p}$ is usually considered as the wavefunction (WF) renormalization. 
It modifies the potential in the next order of perturbation theory \cite{Choi:2019ixb}. 
Notice, however, that the renormalization being proportional to
$g_{\alpha k} g_{\beta k}^*$, 
is different for different neutrino species and therefore 
contributes to oscillations. It can be important when perturbation  
theory starts to break down. 

Let us consider this in some detail. It is simpler to use the diagonalized 
(effective mass-squared) basis. The equation of motion for $i-$th neutrino component can be parameterized as 
\begin{equation}
\left(\slashed{p} + \slashed{p}\Sigma_1^i + 
\slashed{k}\Sigma_2^i
\right) u_{Li} = 0\,,
\label{eq:eqmotion}
\end{equation}
where $\Sigma_{1,2}^i$ are the results of computation of the scattering amplitude.
Here we assume that $\langle \phi \rangle = 0$, so that any extra contribution 
to the equation of motion from the interaction terms in (\ref{eq:Lag}) is absent.
From this equation, we obtain
\begin{equation}
\slashed{p}\,u_{Li} = - 
\frac{\slashed{k}\Sigma_2^i}{1 + \Sigma_1^i}\, u_{Li}\,.
\label{eq:uslashed}
\end{equation}
The amplitude of forward  scattering is given by
\begin{equation}
A_i = \bar{u}_{Li} \left(\slashed{p}\Sigma_1^i + 
\slashed{k}\Sigma_2^i \right) u_{Li}\,.
\label{eq:amplitude}
\end{equation}
Inserting here the expression for $\slashed{p}u_{Li}$ from
(\ref{eq:uslashed}), we obtain 
\begin{equation}
A_i = \bar{u}_{Li}\slashed{k}u_{Li} \Sigma_2^i 
\left(1   - \frac{\Sigma_1^i}{1 + \Sigma_1^i}
\right)\,.
\label{eq:amplitude2}
\end{equation}
For non-relativistic $\phi$, $\slashed{k} \approx 
m_\phi \gamma^0$. Eq.\,(\ref{eq:amplitude2}) gives the potential with the correction due to $\slashed{p}$ term included, 
\begin{equation}
\tilde{V}^i = \frac{V^i}{1 + \Sigma_1^i}\,.
\label{eq:potentials}
\end{equation}
The factor $1/{(1 + \Sigma_1^i})$ is nothing but the effect 
of the WF renormalization. 
It is the next order correction in $g^2$ to $V$. In the perturbative regime, $\Sigma_1^i \ll 1$ and $\tilde{V}^i \approx 
V^i$. 

Each of the factors $\Sigma_{1,2}^i$ has contributions
from the $s-$ and $u-$ channel: 
$$
\Sigma_1^i = \Sigma_{1s}^i+ \Sigma_{1u}^i, \, \, \, 
\Sigma_2^i = \Sigma_{2s}^i+ \Sigma_{2u}^i\,.
$$
Furthermore, as can be seen from the expression for the propagator (\ref{eq:propagator}), the relations 
\begin{equation}
\Sigma_{1s}^i = \Sigma_{2s}^i, \, \, \, 
\Sigma_{1u}^i = - \Sigma_{2u}^i\,,
\label{eq:usrelations}
\end{equation}
exist. In the lowest order, the potential equals 
\begin{equation}
V^i = m_\phi (\Sigma_{2s}^i + \Sigma_{2u}^i)\,.
\label{eq:potcorrection}
\end{equation}
Let us consider the correction due to WF renormalization for some specific cases: 

\begin{enumerate}

\item For $\epsilon = -1$ (resonance $s-$contribution only), we have
$\Sigma_{1u}^i = \Sigma_{2u}^i = 0$,  and from  
(\ref{eq:usrelations}, \ref{eq:potcorrection}), the perturbativity condition $\Sigma_{1}^i=\Sigma_{2}^i=V^i/m_\phi$. Therefore,
$$
\tilde{V}^i = \frac{V^i m_\phi}{m_\phi +  V^i}\,.
$$
Consequently,
\begin{equation}
\tilde{V} = 
\left\{
\begin{matrix}
V, ~ ~ m_\phi \gg  V\,\,,  \\
m_\phi, ~ ~ m_\phi \ll  V\,\,.
\end{matrix}
\right.
\label{eq:vvtilde}
\end{equation}
The case $m_\phi \ll  V$, however, corresponds to non-perturbative 
situations.  The critical value is given by $V = m_\phi$.
The potential equals
$$
V = \frac{\lambda \bar{n}}{2Em_\phi - m_\chi^2}
=  \frac{\Delta m^2_{\rm atm} m_\phi}{2Em_\phi - m_\chi^2}\,.
$$
The equality $V(E_p) = m_\phi$ determines the minimal energy $E_p$,
down to which perturbativity holds. This reads
$$
E_p \simeq \Delta m^2_{\rm atm}/2m_\phi\,.
$$
For $m_\phi =10^{-9}$ eV (which, according to Fig.\,\ref{fig:bounds}, 
can be achieved for $m_\chi=3\cdot 10^{-4}\,$eV),  we find $E_p \simeq 1\,$MeV. 
In this case, perturbativity conditions are satisfied for $E\gtrsim 1\,$MeV. 
This is the relevant energy scale for neutrino oscillation experiments. 
For lower energy, perturbativity is broken, and the results derived (\ref{eq:PotPhi}) cannot be used, 
although some features like energy dependence of the refractive mass may still hold.

\item For $\epsilon = 1$ (no resonance), we have
$\Sigma_{1s}^i = \Sigma_{2s}^i = 0$ 
and $\Sigma_{1u}^i = - \Sigma_{2u}^i = - V/m_\phi$
and the situation is similar to that in the first case  with $\tilde{V}^i = V^i m_\phi/(m_\phi-V^i)$.

\item  For $\epsilon = 0$, explicit computations give
$$
\Sigma_{1}^i /\Sigma_{2}^i =  \frac{m_\chi^2}{2Em_\phi}\,,
$$
therefore
\begin{equation}
\Sigma_1 = \frac{V}{m_\phi}\frac{m_\chi^2}{2 m_\phi E}
= \frac{V}{m_\phi}\frac{E_R}{E}\,,
\label{eq:EpertEp0}
\end{equation}
so above the resonance, the correction is suppressed by $E_R/E$. 
Consequently, the perturbativity energy can be obtained from $\Sigma_1(E_p)=1$. According to (\ref{eq:EpertEp0}), 
$$
E_p = \frac{m_\chi}{2m_\phi}\sqrt{\Delta m^2_{\rm atm}}\,\,.
$$
For $m_\chi =3 \cdot 10^{-4}$ eV and
$m_\phi =10^{-9}\,$eV,  we find $E_p = 1.5 \cdot 10^4\,$eV, while $E_R\simeq 50\,$eV. 
The relation between $E_p$ and $E_R$ is given by
$$
E_p = E_R \frac{\sqrt{\Delta m^2_{\rm atm}}}{m_\chi}\,.
$$
Therefore, perturbativity holds down to resonance energy only 
if $m_\chi>\sqrt{\Delta m^2_{\rm atm}}= 0.05\,$eV.  
Below resonance, perturbativity would require $\Delta m^2_{\rm atm} \ll m_\chi^2$, 
which is not satisfied for the observed values of neutrino masses. 

\end{enumerate}
%%%%%%%%%%%%%%%%%%%%%%%%%%%%%%%%%%%%%%%%%%%%%%

%%%%%%%%%%%%%%%%%%%%%%%%%%%%%%%%%%%%%%%%%%
\subsection{High order corrections and resummation}
%%%%%%%%%%%%%%%%%%%%%%%%%%%%%%%%%%%%%%%%%%%%%%%
\begin{figure}[!t]
\includegraphics[width=0.9\textwidth]{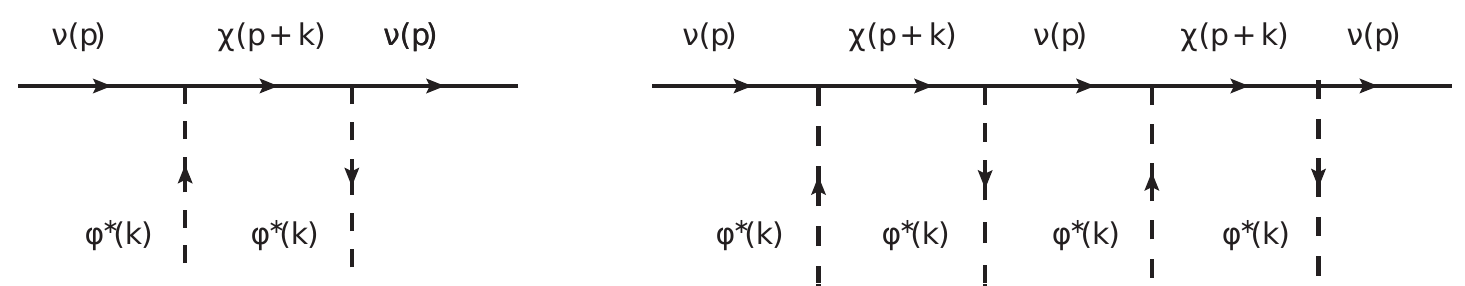}
\caption{ $s-$ channel $\nu-\phi$ interaction processes at the tree level (left) 
and at the next order with two $\phi$ being absorbed and emitted (right). 
The $u-$channel amplitude corresponds to diagrams with intersecting scalar lines.}
\label{fig:Nu-Phi-scattering}
\end{figure} 
%%%%%%%%%%%%%%%%%%%%%%%%%%%%%%%%%%%%%%%%%%%%%%%%%%%%%%%%%%%%%%%%%%

Explanation of values of neutrino masses implies that  $r_\chi \gg d$  in contrast to
the standard case ($r_W \ll d$) which may testify that
neutrino interactions with many $\phi$ can be
important (especially if both $\phi$ and $\phi^*$ are around).
So, one should consider processes with more than one
$\phi$ being absorbed and emitted: 
$$
\nu \phi \rightarrow  \nu \phi ,  \, \, \,
\nu \phi  \phi \rightarrow  \nu \phi \phi ,  \, \, \,
\nu \phi \phi \phi \rightarrow  \nu \phi \phi \phi ,\, \cdots\,
$$
as shown in Fig.\,\ref{fig:Nu-Phi-scattering}.
The corresponding diagrams contain alternate
propagators of $\chi$ and $\nu$ with $\phi$ vertices.

Let us consider, for definiteness, the case $\epsilon = -1$ when the $s-$channel contributes only.
In the presence of two $\phi$ interactions $(\nu \phi  \phi \rightarrow  \nu \phi \phi )$ , 
the amplitude picks up an additional factor
\begin{equation}
\zeta = \lambda \frac{n_\phi}{m_\phi}\frac{\slashed{p} \slashed{p}_\chi}{p^2 (p_\chi^2 - m_\chi^2)}
= \Delta m_{\rm atm}^2 \frac{\slashed{p} (\slashed{p} + \slashed{k})}{p^2 (p_\chi^2 - m_\chi^2)}\,,
\label{eq:defz}
\end{equation}
which should act eventually on the neutrino spinor $u_L$ to the right. Permuting $\slashed{p}$ and $\slashed{k}$,
and using Eq.\,(\ref{eq:uslashed}), we obtain
\begin{equation}
\zeta
= \Delta m_{\rm atm}^2 \frac{p^2 + 2(pk) - m_\phi \tilde{V}}{p^2 (p_\chi^2 - m_\chi^2)}.
\label{eq:defz2}
\end{equation}
(The corrected potential $\tilde{V}$ is determined in eq. (\ref{eq:potentials}).)  
The dispersion relations reads $p^2 = 2VE  + V^2 \approx 2VE$. The other term $2(pk) = 2E m_\phi$, and consequently, 
\begin{equation}
\zeta \approx \frac{\Delta m_{\rm atm}^2}{p_\chi^2 - m_\chi^2} = \frac{V}{m_\phi} = \Sigma_1\,.
\label{eq:defz3}
\end{equation}
Thus, the perturbativity condition $\zeta \ll 1$ coincides  with the condition of smallness of
WF renormalization (see (\ref{eq:potentials})). This is not accidental 
since the neutrino WF renormalization is described through absorption and emission of multiple $\phi$, 
and therefore is described by the same diagram as Fig.\,\ref{fig:Nu-Phi-scattering} (right). 
Similarly, one can show the equivalence of perturbation conditions for $\epsilon = 1$ and $\epsilon  = 0$.

Formally, resummation of the diagram with various number of insertions gives
\begin{equation}
V_{\rm resum} = \frac{V}{1 + \zeta}\,,
\label{eq:defz3}
\end{equation}
which can lead to an additional suppression of the potential.
As we saw, perturbativity holds above resonance.
Below resonance $\zeta \simeq  \Delta m_{\rm atm}^2/ m_\chi^2 \gg 1$ (for $\epsilon = -1$),
so the result cannot be estimated reliably. 
However, below resonance, $\zeta$ does not depend on $E$ and therefore resummation should not
produce an energy dependence of $V$ at least for $\epsilon\neq 0$.  Hence,
one expects a decrease of the effective mass-squared with $E$ simply due to multiplication of the potential by $(2E)$.

The convergence of this series of interactions can be tested by checking whether 
the $n-$th term of the series diverges. For a real field, the $n-$th term in the series gets 
a contribution of $n!$ from the different $\phi$ permutations. This cancels with 
the factor of $(1/n!)$ from the exponential, giving $(1/n!)\,n!\zeta^n=\zeta^n$. 
This series sum leads to $1/(1+\zeta)$, which is convergent. For a complex field, the number of permutations 
allowed are $(n/2)!$ for $\phi$ and $\phi^*$ respectively. 
This leads to a factor of $(1/n!)\,(n/2)! (n/2)! \zeta^n$ in the $n-$th term. 
Therefore, for a complex $\phi$, the series convergence is better due to an extra suppression by $n$.

%%%%%%%%%%%%%%%%%%%%%%%%%%%%%%%%%%%%%%%%%%%%%%%%%%%%%%%%%%%%%%%%%%%%
\subsection{The $\nu- \chi$ mixing due to scattering}
%%%%%%%%%%%%%%%%%%%%%%%%%%%%%%%%%%%%%%%%%%%%%%%%%%%%%%%%%%%%%%%%%%

Interaction (\ref{eq:Lag}) in a medium composed of the cold gas of scalars also produces
$\nu_L - \chi^c_L$ mixing. The mixing is generated by off-diagonal potential in the 
space of states $(\nu, \chi^c_L)$ which is proportional to 
the amplitude of forward off-diagonal scattering. The amplitude in the unit of length equals
\begin{equation}
A_{\alpha k}(x) =   g_{\alpha k} \langle \chi^c_{k L} | \chi_{k R} \rangle    \sum_{j = 1 \div n_\phi} 
\langle \chi_{k R}(p_\chi) |\,\bar{\chi}_k (x) \nu_\alpha (x) \phi^*(x) \,  
| \nu_L (p_\nu) \phi^*(p_j) \rangle\, ,
\label{eq:nu-chimix}
\end{equation}
where summation proceeds over scatterers on the unit of length. 
Taking into account that $\langle \chi^c_{k L} | \chi_{k R} \rangle = m_{\chi k}/2E_\chi$ 
the explicit computations give the off-diagonal potential  
\begin{equation}
V_{\alpha k}(x) = \frac{m_{\chi k}}{2E_\chi} \frac{g_{\alpha k}}{\sqrt{2 m_\phi}}
e^{i (p_\chi - p_\nu) x} \sum_{j = 1 \div n_\phi} e^{- i p_j x} .  
\label{eq:offdpot}
\end{equation}
For ${\bf x}=0$ and non-relativistic scalars,  $E_j \simeq m_\phi + m_\phi v_j^2/2$, hence
the sum in Eq.(\ref{eq:offdpot}) equals 
for random phases
\begin{equation}
e^{-i m_\phi t} \sum_{j = 1 \div n_\phi} e^{-i m_\phi v_j^2 t/2}\simeq  e^{-i m_\phi t + i\phi'}  \sqrt{n_\phi},  
\label{eq:sumph}
\end{equation}
where $\phi'$ is some residual phase.

Consequently, the off-diagonal potential equals
\begin{equation}
V_{\alpha k}(x) =  e^{-i m_\phi t + i\phi'}\,  g_{\alpha k} \frac{m_{\chi k}}{2E_\chi}  
\sqrt{\frac{n_\phi}{2\,m_\phi}}\, ,  
\end{equation}
where we have taken $E_\nu = E_\chi$ which removes the phase in eq. (\ref{eq:offdpot}) related to $\nu$ and $\chi$. \\

Now the Hamiltonian of complete system which includes
active neutrinos and singlets $\chi_k$ in the basis
$(\nu_\alpha, \chi_k^c)_L$ can be written in the block form as
\begin{equation}
{\bf H} \approx 
\frac{1}{2E}
\begin{bmatrix}
\tilde {m}_{\alpha \beta}^2   &  e^{-im_\phi t + i\phi'}
g_{\alpha k} m_{\chi k}\sqrt{n_\phi/2 m_\phi} \\
e^{im_\phi t - i\phi'}
g_{\alpha k} m_{\chi k}\sqrt{n_\phi/2 m_\phi} &
m_{\chi k}^2 + \tilde {m}_{\chi k}^2
\end{bmatrix},
\label{eq:complete}
\end{equation}
where $\tilde {m}_{\chi k}$ is refractive
contribution to the mass squared of $\chi$. Parameters $m_{\chi k}$, $\phi'$
can be selected in such a way that
feedback of the $\nu - \chi$ mixing on the active neutrino oscillations is small.
We consider phenomenological consequences
of this mixing in sect.\,\ref{sec:nuchiosc}.

\subsection{Resonance production of $\chi$}
%%%%%%%%%%%%%%%%%%%%%%%%%%%%%%%%%%%%%%%%%%%%%%%%%%%%%%%%%%%%%%%%%%%%%%%%%%%%%

We can also consider the resonant production $\nu \rightarrow \chi$ in the bath of cold scalar particles.
The amplitude of transition $\nu (p) + \bar{\phi} (k) \rightarrow \chi (p_\chi)$ is proportional to
%%%%%%%%
\begin{equation}
g \phi_0\left( \frac{m_\chi}{2 E_\chi} \right)\,\int d x\, e^{i (p + k - p_\chi) x},
\label{eq:nuchitran}
\end{equation}
%%%%%%%%%%5
where the factor $m_\chi / 2 E_\chi$  follows from angular momentum conservation
in the scalar field, which produces the flip $\chi_R \rightarrow \chi_L$.
The amplitude of scalar field $\phi_0$ is given by number density of $\phi$.

Integration over infinite space-time volume leads to a delta-function signifying
exact energy-momentum conservation $p + k = p_\chi$. For $k = (m_\phi, 0)$,
in the limit $m_\chi \gg m_\phi$, energy-momentum conservation  gives $E = p = m_\chi^2/2m_\phi$,
which is exactly the resonance energy $E_R$, as defined in (\ref{eq:resen}).
The uncertainty in energy $\delta E$ related to $\chi$ decay,  $\chi \rightarrow \bar{\phi} + \nu$,
is negligible: it is given by the width $\Gamma_{\chi, k} = g^2 m_\chi/ 8 \pi$.
For instance, for $m_\chi = 10^{-5}$ eV and $g = 10^{-12}$, we find $\Gamma_\chi = 4 \cdot 10^{-19}$ eV and
 $\delta E = \Gamma_\chi  m_\chi /E \sim 10^{-30}$ eV.

Much bigger uncertainty arises from non-zero velocities of $\phi$ particles.
The energy and momentum conservation laws give
\begin{eqnarray}
\label{eq:enercons1}
p_\chi & = & p  \pm m_\phi v\, ,\\
p_\chi & + & \frac{m_\chi^2}{2 p_\chi}  =
p + m_\phi + \frac{m_\phi v^2}{2}\, .
\label{eq:enercons}
\end{eqnarray}
Here $\pm$ in the Eq.\,(\ref{eq:enercons1}) reflects interval of possible changes of $\phi-$momentum.
From Eq. (\ref{eq:enercons}), we find
\begin{equation}
E_R \approx p_R \approx  \frac{m_\chi^2}{ 2 m_\phi (1 \pm v ) }\, .
\end{equation}
That is,  the resonance peak acquires the width
$$
\Delta E \simeq  2E_R v\,.
$$

\begin{figure}[!t]
\includegraphics[width=0.7\textwidth]{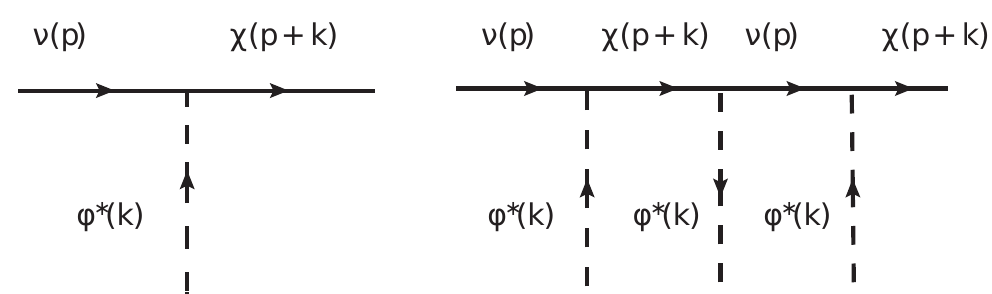}~
\caption{Left: $\nu-\chi$ transition at tree-level. Right:  $\nu-\chi$ transition with additional $\phi$ interactions.}
\label{fig:Nu-Chi-transition}
\end{figure} 

Furthermore, the transition $\nu \rightarrow \chi$ occurs in the $\phi-$medium which can be accounted in the lowest order by diagrams with two additional $\phi$ interactions  in $\nu$ and $\chi$ lines, as shown in Fig.\,\ref{fig:Nu-Chi-transition}. This corresponds to renormalization of the
$\nu-$ and $\chi-$ wave functions. The resulting two diagrams are identical giving corrections $\sim 2\Sigma_1$, where the factor of 2 is nothing but the Bose enhancement.

Finally, one should take into account the energy uncertainty in the propagating neutrinos which is determined by the process of their production.
The uncertainty due to finite space-time integration in (\ref{eq:nuchitran}), given by inverse of baseline $L$ of experiment,  $\sigma_p \sim 1/L$, is usually much smaller.

Thus, only a very small fraction of neutrinos $\simeq (2\cdot 10^{-3} - 10^{-2})$ with energies close to the resonance energy can be converted into $\chi$. Far from the resonance $(E\gg E_R$ and $E\ll E_R)$, $\chi$ is strongly virtual and can exist for a very short period of time $\sim 1/E\sim 10^{-20}\,$s. 

%%%%%%%%%%%%%%%%%%%%%%%%%%%%%%%%%%%%%%%%%%%%%%%%%%%%%%%%%%%%%%%%%%%%%%%%%%%%%%%%%%%%%%%%%%%%%%%%%%%%%%%%%%%%%%
\section{Refraction in the classical field}
\label{sec:RefMassField}
%%%%%%%%%%%%%%%%%%%%%%%%%%%%%%%%%%%%%%%%%%%%%%%%%%%%%%%%%%%%%%%%%%%%%%%%%%%%%%%%%%

Finding refractive mass in a cold gas
via computations of the amplitudes of scattering on individual scatterers and summation over scatterers
show some potential problems
(unusual relation of scales, perturbativity, etc.). This may demand some other approach
to computations. 

\subsection{State of the scalar field}
%%%%%%%%%%%%%%%%%%%%%%%%%%%%%%%%%%%%%%%%%%%%%%%%%%%%%%%%%%%%%%%%%%%%%%%%%%%%%%%%%%%%%%%%%%%%%%%%%%

To explain neutrino masses, one should have a very large number density of $\phi$. 
If the occupation number of $\phi$ in the volume determined by the de Broglie wavelength is much bigger than 1, that is,
 $\lambda_{dB}^3 n_\phi \gg 1$, a system of scalar particles can  be treated 
as classical field  $\phi_c$ (classicality condition)
\footnote{The classicality condition can also be written as $\lambda_{dB}\gg d$.}. 
Using that $n_\phi = \rho_\phi/m_\phi$,
we obtain from the classicality condition,
\begin{equation}
m_\phi \ll 2\pi \left[\frac{\rho_\phi}{2\pi v^3} \right]^{1/4}.
\label{eq;class}
\end{equation}
For $v_{\rm vir} = 10^{-3}$ and $\rho_\phi = \rho_{\odot} = 0.3$ GeV/cm$^3$, the equation
(\ref{eq;class}) gives  $m_\phi \ll 30$ eV.  Then, for $E_R = 0.01\,$MeV, Eq.\,(\ref{eq:resen}) would lead to $m_\chi \ll 770\,$eV.

Such a scalar field can appear as an expectation value of the field operator in a coherent state of particles:
\begin{equation}
\phi_c (x) = \langle \phi_{\rm coh} |\hat{\phi}(x) |\phi_{\rm coh} \rangle\,,
\label{eq:expval}
\end{equation}
where the coherent state $|\phi_{\rm coh} \rangle$ and its complex 
conjugate $|\bar{\phi}_{\rm coh} \rangle$ are defined in the Appendix\,\ref{sec:Appendix}. These states could be formed in the early Universe
via the misalignment mechanism.

The most general state is given by a linear combination 
of $|{\phi}_{\rm coh} \rangle$ and $|\bar{\phi}_{\rm coh} \rangle$,
\begin{equation}
|{\phi}_{\rm coh}^{\rm tot} \rangle = \cos\alpha |{\phi}_{\rm coh} \rangle + \sin \alpha |\bar{\phi}_{\rm coh} \rangle \,.
\label{eq:totalst}
\end{equation}
The corresponding field can be written as 
\begin{equation}
\phi_c^{\rm tot}(x) = \langle{\phi}_{\rm coh}^{\rm tot}|\hat{\phi}(x) |{\phi}_{\rm coh}^{\rm tot} \rangle
= \int \frac{d^3k}{ (2\pi)^3} \frac{1}{\sqrt{2 E_k}}
\left[f_a^{\rm tot}(k) e^{-i k x} + f_b^{\rm tot}(k) e^{i k x} \right]\,.
\label{eq:phiccom}
\end{equation}
$\phi_c^{\rm tot}(x)$ can also be parameterized as (for a detailed derivation, see Appendix\,\ref{sec:Appendix})  
%%%%%%%%%%
\begin{equation}
\phi_c^{\rm tot}(x) =   F(x) e^{i\Phi} .
\label{eq:paramp}
\end{equation}
%%%%%%%%
For the Hermitian conjugate, we have similarly 
\begin{equation}
\phi_c^{{\rm tot}~\dagger}(x) =  \bar{F}(x) e^{-i \bar{\Phi}}. 
\label{eq:parampdag}
\end{equation}
The factors  $F^2$ and $\bar{F}^2$ are related to the corresponding contributions
of $\phi$ to the energy density in the Universe:
\begin{equation}
F^2  = \frac{\rho_\phi}{m_{\phi}^2}, ~~~ \bar{F}^2  = \frac{\rho_{\bar \phi}}{m_{\phi}^2}\,.
\label{eq:Fv0}
\end{equation}
For a real field, we have $F(x)=\bar{F}(x) $ and $\tan\Phi=0$.

%%%%%%%%%%%%%
\subsection{Decoherence and strength of the scalar field}
%%%%%%%%%%%%%%%%%%%%%%%%%%%%%%%%%%%%%%%%%%%%%%%%%%%%%%%%%%

Even if at the moment of creation, the coherent state has
$k \simeq 0$ and $E \simeq m_\phi$, in the course of
DM halo formation (virialization), due to gravitational
interactions, $\phi$ particles acquire a velocity distribution dispersion width $v_{\rm vir}$.
In our Galaxy, the virialized velocities
$v_{\rm vir} \simeq 10^{-3}$, and corresponding momenta
$k = m_\phi v_{\rm vir}$. This determines the
distributions functions of $\phi$ particles, described through $f(k)$ in (\ref{eq:phiccom}). Due to dispersion of velocity,
different $\phi$ acquire different momenta and energies, and consequently phases, which lead to decoherence.
This appears as a suppression of the coherent component of the fields due
to integration over momenta in (\ref{eq:phiccom}).

For a given spatial point, the phase difference between different modes  equals
\begin{equation}
\Delta \phi =  \Delta E t \simeq E_{\rm vir} t = \frac{1}{2}
m_\phi v_{\rm vir}^2 t\,.
\end{equation}
When $\Delta \phi \sim 2\pi$, coherence is lost.
This gives for coherence time
\begin{equation}
t_{\rm coh} \simeq \frac{4\pi}{m_\phi v_{\rm vir}^2} =
8 \cdot 10^{-9}\,  {\rm s}  \left(\frac{\rm eV}{m_\phi}\right).
\label{eq:tdecoh}
\end{equation}
In reality, the time is bigger since $v_{\rm vir}$ increases in the
process of virialization. For $m_\phi = 10^{-12}\,{\rm eV}, \, 10^{-16}\,{\rm eV},\,
10^{-20}\,{\rm eV}$, we obtain
$t_{\rm coh} = 8 \cdot 10^{3}\,{\rm s},\,\, 8 \cdot 10^{7}\,{\rm s}, \,\, 8 \cdot 10^{11}\,{\rm s}$
respectively. This implies that the field decoheres already during virialization.
One can also compute the distance travelled by $\phi$ before decoherence takes over:
\begin{equation}
x_{\rm coh} \simeq \frac{4\pi}{m_\phi v_{\rm vir}} =
5 \cdot 10^{8} \, {\rm cm} \,\left( \frac{ 5 \cdot 10^{-10}~{\rm eV}}{m_\phi}\right).
\end{equation}
Thus, only for very tiny masses of $\phi$, coherence can be maintained over cosmological time.

In fact, for non-relativistic $\phi$, the scalar field does not vanish
completely due to decoherence. The ``residual" classical field will exist even for $t \rightarrow \infty$.
Let us consider a unit volume at some spatial point and substitute the
integration over momenta in (\ref{eq:phiccom}) by summation:
\begin{equation}
\phi_c^{\rm tot}(x)
\approx \frac{1}{\sqrt{2 m_\phi}} \sum_k
\left[f_{a k} e^{-i k x} + f_{b k} e^{i k x} \right].
\label{eq:phicsum2}
\end{equation}
Taking ${\bf x} =0$ and $E_k \simeq m_\phi + m_\phi v_k^2/2$, we obtain
\begin{equation}
\phi_c^{\rm tot}(x)
= \frac{1}{\sqrt{2 m_\phi}}
\left[e^{-i m_\phi t} \sum_k e^{-i m_\phi v_k^2 t/2 }
+ e^{i m_\phi t} \sum_k e^{i m_\phi v_k^2 t/2} \right].
\label{eq:phicsum3}
\end{equation}
For random phases in the exponents under sums,
we have
$\sum_k e^{i m_\phi v_k^2 t/2 } = \sum_k e^{-i m_\phi v_k^2 t/2 } \approx \sqrt{n}$, and therefore
\begin{equation}
\phi_c^{\rm tot}(t) \sim \sqrt{\frac{n_\phi}{m_\phi}} \cos m_\phi t \,.
\label{eq:phicsum4}
\end{equation}
The amplitude matches the field strength used in the cold gas computations. In addition, time variations of the field appear which have been studied extensively in a number of papers~\cite{Berlin:2016woy,Krnjaic:2017zlz,Brdar:2017kbt,Capozzi:2018bps,Dev:2020kgz,
Losada:2021bxx,Huang:2021kam,Chun:2021ief,Dev:2022bae,Huang:2022wmz,Davoudiasl:2023uiq,Losada:2023zap,Gherghetta:2023myo}.

%%%%%%%%%%%%%%%%%%%%%%%%%%%%%%%%%%%%%%%%%%%%%%%%%
\subsection{Hamiltonian of propagation}
%%%%%%%%%%%%%%%%%%%%%%%%%%%%%%%%%%%%%%%%%%%%%%

The interaction (\ref{eq:Lag}) in the presence of non-zero expectation value of $\phi$
generate mass terms in the same way as the VEV of Higgs field generates masses in the Standard Model 
\begin{equation}
%%\langle \phi_{\rm coh} |\mathcal{L}|\phi_{\rm coh} \rangle =
\mathcal{L} \supset
%%\sum_k g_{\alpha k} \bar{\chi}_{k R}\nu_{\alpha L} \phi^*_q   +
%%\sum_k g_{\alpha k}^* \bar{\nu}_{\alpha L} \chi_{k R} \phi_q +
\sum_k g_{\alpha k} {\phi}_c^{{\rm tot}~\dagger}  \bar{\chi}_{k R}\nu_{\alpha L}
+ \sum_k g_{\alpha k}^* \phi_c^{\rm tot} \bar{\nu}_{\alpha L} \chi_{k R}\,.
\label{eq:masst}
\end{equation}
Notice that our consideration of evolution differs
from that in \cite{Chun:2021ief}.
The resulting mass matrix in the basis of
3 flavor  neutrino states  and two mediators with definite Majorana masses:
$(\nu_f, \chi_L)^T =  (\nu_e, \nu_\mu, \nu_\tau, \chi_1, \chi_2)^T$,  where $\chi_L \equiv (\chi_R)^c$,
can be written  as
\begin{equation}
{\bf M} =
\begin{bmatrix}
0    ~~~~   & g_{\alpha k} {\phi}_c^{{\rm tot}~ \dagger}   \\
g_{\alpha k} {\phi}_c^{{\rm tot}~ \dagger}    ~~~~    & {\rm diag} (m_{\chi k})
\end{bmatrix}\,,
~~~ \alpha = e, \mu, \tau, ~~~k = 1, 2.
\label{eq:massm1}
\end{equation}
The Hamiltonian of evolution is given by
\begin{equation}
{\bf H} \approx \frac{1}{2E} {\bf M} {\bf M}^\dagger =
\frac{1}{2E}
\begin{bmatrix}
\bar{F}^2 \sum_k  g_{\alpha k}  g_{\beta k}^*  ~~~  & g_{\alpha k} \bar{F} m_{\chi k} e^{- i \bar{\Phi}}  \\
g_{\alpha k}^* \bar{F}^* m_{\chi k} e^{i \bar{\Phi}}  ~~~ &   \bar{M}_\chi^2
\end{bmatrix},
\label{eq:hamnu}
\end{equation}
where a common term proportional to the identity matrix is omitted and $\bar{F},\,\bar{\Phi}$ are defined in (\ref{eq:paramp}). 
Here
\begin{equation}
\bar{M}_\chi^2 =
\begin{bmatrix}
\bar{F}^2  \sum_\alpha  |g_{\alpha 1}|^2   + m_{\chi 1}^2
~~~ & \bar{F}^2 \sum_\alpha  g_{\alpha 1}  g_{\alpha 2}^*    \\
\bar{F}^2 \sum_\alpha  g_{\alpha 2}  g_{\alpha 1}^*
~~~  &   \bar{F}^2 \sum_\alpha  |g_{\alpha 2}|^2  + m_{\chi 2}^2
\end{bmatrix}.
\end{equation}
%%%%%%%%%%
%%%%%%%%%%
For antiparticles (basis $(\bar{\nu}, \bar{\chi})^T$), the mass matrix equals according to
(\ref{eq:masst})
\begin{equation}
\bar{\bf M} =
\begin{bmatrix}
0    ~~~~   & g_{\alpha k}^* {\phi}_c^{\rm tot}   \\
g_{\alpha k}^* {\phi}_c^{\rm tot}    ~~~~    & {\rm diag} (m_{\chi k})
\end{bmatrix}  ~~~~~ \alpha = e, \mu, \tau, ~~~k = 1, 2.
\label{eq:massm2}
\end{equation}
Consequently,  the Hamiltonian can be obtained from (\ref{eq:hamnu}) by the following substitution:
$\bar{F} \rightarrow F\,\,$,
$\bar{\Phi} \rightarrow {- \Phi}$.

The Hamiltonian of evolution  (\ref{eq:hamnu})  coincides
with  the Hamiltonian (\ref{eq:complete}) computed for
cold gas at energies above the resonance,  if $\Phi = m_\phi t - \phi'$.
The $3\times 3$ block of active neutrinos in the Hamiltonian (\ref{eq:hamnu}), as well as the off-diagonal mixing block
have the same form  as the matrix $\tilde{m}_{\rm asy}^2$ produced by refraction on a cold $\phi$ gas
(\ref{eq:minf}) at high neutrino energies. We call $\bar{F}^2 \sum_k  g_{\alpha k}  g_{\beta k}^*$ the refractive mass-squared in this case. We discuss in details two major features of this Hamiltonian:
(1) energy and space - time dependence of  neutrino effective mass matrix, and  (2)
$\nu - \chi$ mixing  and oscillations.

%%%%%%%%%%%%%%%%%%%%%%%%%%%%%5
\subsection{ Energy and space - time dependence of  neutrino effective mass matrix}
%%%%%%%%%%%%%%%%%%%%%%%5
Refractive mass depends on the neutrino energy,  in particular, it 
decreases with energy below the resonance. In contrast, 
the mass (\ref{eq:hamnu}) generated by the expectation value of the coherent scalar field  on the first sight,   
has no resonance
and does not depend on energy. Such a behavior would correspond
to $E_R \rightarrow 0$ and $m_\chi = 0$. The controversy is resolved in the following way.

The coherent field $\phi$ has time dependence and therefore can be considered as
a wave with frequency $\sim m_\phi$.
Therefore we can describe  the process  $\nu + \phi \rightarrow  \chi$, as the  
$\nu \rightarrow \chi$ transition with the absorption of the $\phi$ wave. 
%%(To have more common picture one can go to moving reference
%%frame so that  $\phi$  has the form of usual $t$ and $x-$ dependent wave.)
%%This is similar to  excitation of atomic level
%%under absorption of EM wave.
The transition occurs when  frequency $m_\phi$ matches difference of energies of $\nu$ and 
final state $\chi$ and therefore has a resonance character. 
Below resonance,  there is no on-shell $\chi$ production,  therefore $\chi$ appears as a  virtual particle,  
and the amplitude of its appearance decreases with increase of virtuality. Similarly, above the resonance, 
the appearance of $\chi$ decreases as $1/E$.

In fact, the wave component of the field $\phi(t, x)$ can be immediately treated as
particle and therefore refraction effect can be computed with $\chi$ propagator 
 in the same way as in a cold gas.

%%Studying matter potential  we are interested in $\nu \rightarrow \nu$ transition
%%in time dependent external field. 
%%This can be done computing neutrino propagator in the external field 
%%[high order corrections - Schwinger - Dyson]

The scalar field depends on time, being proportional to $\bar{F}(x)$.  
%%[[but such ??? a dependence also 
%%appears in refraction in the cold gas]]. 
Additional dependence on time is in the phase $\Phi$. 
%%in addition to dependence of $n_\phi (t)$.
For a real field, in the non-relativistic approximation $(k \simeq 0)$, the scalar mass squared goes as
$4\bar{F}^2 \cos^2 m_\phi t$~\cite{Berlin:2016woy}.
%${\phi}_0$ or $\bar{\phi}_0$ vanish, that is, if  
Notice that $\bar{F}^2$ (${F}^2$) appears as a prefactor in the active neutrino block of the Hamiltonian. 
therefore its time variations do not change mixing of active neutrinos, 
but do change the effective masses squared. The scale of masses of neutrinos and antineutrinos,
given by $\bar{F}^2$ and $F^2$, are the  same for a real scalar field but 
can be different for a complex scalar field. This can imply that the refractive masses
for neutrinos and antineutrinos can be different.

For very small $m_\phi$, the period of time variations, $(T_{\phi} \propto 1/m_\phi)$,
is very long - much longer than time of neutrino propagation
in the oscillation setup. So, for description of oscillations
one can take constant $\bar{F}^2(t_0)$ and  $F^2 (t_0)$
in a given moment $t_0$. This may not be true for cosmic neutrinos (SN or high energy neutrinos).
However, $T_\phi$ can be smaller or comparable to the duration of neutrino experiments
(several years). So, one can observe changes of extracted
$\Delta m^2$ in the course of experiment. Non-observations of
such changes put bounds on parameters of the
field~\cite{Berlin:2016woy,Krnjaic:2017zlz,Brdar:2017kbt,Dev:2020kgz}.

%%%%%%%%%%%%%%%%%%%%%%%%%%%%%%%%%%%%%%%%%%%%%%%%%%%%%%%%%%%%%%%%%%%%%%%%%%%%%%%%%%%%%
\subsection{The $\nu - \chi$ mixing and oscillations}
\label{sec:nuchiosc}
%%%%%%%%%%%%%%%%%%%%%%%%%%%%%%%%%%%%%%%%%%%%%%%%%%%%%%%%%%%%%%%%%%%%%%%%%%%%

According to the Hamiltonian in (\ref{eq:hamnu}), there is  $\nu_\alpha - \chi_k$ mixing, 
and therefore $\nu_\alpha - \chi_k$ oscillations. A convenient way to study 
this is to partially diagonalize the effective mass matrix (\ref{eq:massm1}). 
For definiteness, we will consider mixing of the neutrinos.

The off-diagonal submatrix of (\ref{eq:massm1}) has the form
\begin{equation}
{\bf m}_{a\chi}  = ||g_{\alpha k} {\phi}_c^{tot~ \dagger}|| =
\bar{F} e^{- i \bar{\Phi}}
\begin{bmatrix}
g_1 ~~~  & 0\\
g_1 ~~~  & g_2\\
g_1 ~~~  & - g_2
\end{bmatrix}.
\label{eq:offdiag}
\end{equation}
It can be diagonalized by a TBM rotation $(U_{TBM})$ of the active components only:
\begin{equation}
U_{TBM} {\bf m}_{a\chi}  =
e^{- i \bar{\Phi}}
\begin{bmatrix}
0 ~~~  & 0\\
m_{a 1} ~~~  & 0\\
0 ~~~  & m_{a 2}
\end{bmatrix},
\label{eq:offdiagdiag}
\end{equation}
where
\begin{equation}
m_{a 1} \equiv  \sqrt{3} g_1\,\bar{F}\,,~~~
m_{a 2} \equiv \sqrt{2} g_2  \bar{F}\, ,
\end{equation}
and  $a$ denotes the rotated basis of active neutrinos given by $\nu_a'=U_{TBM}^T \nu_\alpha$. 
The Majorana mass  matrix  of $\chi$ remains diagonal in the transformation (\ref{eq:offdiagdiag}).
Notice that one active (massless)
state decouples, and in the limit $m_\chi\rightarrow0$, 
the other states $(\nu_{a1},\nu_{a2})$ form two Dirac neutrinos 
with masses $m_{a 1}$, $m_{a 2}$. 
\footnote{
In general, $2 \times 2$ matrix  $a,~ c~ /~ b, ~d$ can be diagonalized
by the left rotations only if $a b = - c d$.
Deviation from TBM does not change results qualitatively.}

After the rotation by (\ref{eq:offdiagdiag}) and decoupling of the $\nu_e'$ state, 
the remaining 4-state system $(\nu_{a1},\nu_{a2},\chi_1,\chi_2)$ 
splits into two independent blocks of $(\nu_{a1}, \chi_1)$ and $(\nu_{a2}, \chi_2)$. 
The $(\nu_{a1}, \chi_1)$ mass matrix is given by
\begin{equation}
{\bf M}_{a 1}  =
\begin{bmatrix}
0 ~~~  & m_{a 1} e^{- i \bar{\Phi}}\\
m_{a 1} e^{- i \bar{\Phi}} ~~~  & m_{\chi 1}
\end{bmatrix}.    
\end{equation}
The corresponding Hamiltonian reads
\begin{equation}
{\bf H}_{a 1} \approx \frac{1}{2E} {\bf M}_{a 1}  {\bf M}_{a 1} ^\dagger =
\frac{1}{2E}  \left( m_{a 1} ^2 {\bf I} +
\begin{bmatrix}
0  & m_{a 1} m_{\chi 1} e^{- i \bar{\Phi}}  \\
m_{a 1} m_{\chi 1} e^{i \bar{\Phi}} &    m_{\chi 1}^2
\end{bmatrix} \right).
\label{eq:hamblock}
\end{equation}

A similar Hamiltonian for $(\nu_{a2}, \chi_2)$ can be obtained by substitution 
$1\leftrightarrow 2$. The two pairs of states evolve independently leading 
to $\nu_{a1} \rightarrow \chi_1$ and  $\nu_{a2} \rightarrow \chi_2$ oscillations.

The complex phase can be removed from the evolution equations 
by rephasing the neutrinos as $\nu_a' \rightarrow e^{i\Phi(t)} \nu_a'$. This leads to the
appearance of an extra term in the Hamiltonian, proportional to the  derivative of $\Phi$: 
\begin{equation}
{\bf H}_{a 1}  \approx
\frac{1}{2E}  \left( m_{a 1} ^2 {\bf I} +
\begin{bmatrix}
0  & m_{a 1} m_{\chi 1}  \\
m_{a 1} m_{\chi 1}   &    m_{\chi 1}^2 - 2 E \dot {\bar{\Phi}}
\end{bmatrix} \right)\,.
\label{eq:hamblock1}
\end{equation}
Here $\dot{\bar{\Phi}} \simeq 0 \div  m_\phi$, where $0$ corresponds 
to symmetric background (real field), while $m_\phi$ is in the case 
of completely asymmetric background (for details, check Appendix\,\ref{sec:Appendix}).

The terms in ${\bf H}_{a i}$ proportional to the unit matrices do not affect flavour evolution
inside the pairs. However, they are important in the whole 5 neutrino system,
in particular, in the relative evolution of the pairs.
Diagonalization of the Hamiltonian (\ref{eq:hamblock1}) gives mass splittings and
mixing: 
\begin{equation}
\Delta m_{a 1} ^2 = 2 \sqrt{(m_{\chi 1}^2 - 2 E \dot {\bar{\Phi}})^2 + m_{a 1} ^2 m_{\chi 1}^2}\,\,,
\label{eq:splitt}
\end{equation}
\begin{equation}
\tan 2\theta_{a 1}  = \frac{2m_{a 1}  m_{\chi 1}}{m_{\chi 1}^2 - 2 E \dot{\bar{\Phi}}}\,.
\label{eq:mixing}
\end{equation}
%%%%%
%%%%%%%
Since $m_{\chi 1}^2\ll m_{a 1}^2 \simeq  \Delta m_{\rm sol}^2$, in (\ref{eq:splitt}), $m_{\chi 1}^2$ can be neglected.
Therefore, for $\dot{\bar{\Phi}}=0$, we get
\begin{equation}
\Delta m_{a 1}^2  \geq 2 \sqrt{\Delta m^2_{\rm sol}} m_{\chi 1}\,,
~~~~\tan 2\theta_{a 1} \simeq\frac{2\sqrt{\Delta m^2_{\rm sol}}}{  m_{\chi 1}}\,.
\end{equation}

For $m_{\chi 1} < 10^{-3}\,$eV, the $\nu - \chi$ oscillations can be relevant 
for neutrinos from astrophysical sources such as the Sun, a core-collapse supernova (SN), 
as well as more distant sources.
Furthermore,  since 
$m_{a 1} \gg m_{\chi 1}$, neutrinos turn out to be pseudo-Dirac particles 
with nearly maximal mixing of the active and sterile components.

Notice that the resonance character of refraction can be seen immediately 
from the Hamiltonian that describes oscillations. 
Indeed,  according to (4.24) for $\dot{\Phi} = m_\phi$ the  
$\chi - \nu$ mixing becomes maximal at $E = m_\chi^2/2 m_\phi$ which is nothing 
but the resonance energy in the potential $V$ due to scattering. So,  the resonance here corresponds to 
oscillations with maximal depth.  
Change of the $\chi - \nu$ mixing parameter with energy corresponds to change of potential due to scattering: below the resonance $\tan 2\theta_{ai}$  converges to a constant value, while above the resonance it decreases as 
$1/E$. 

The total mixing matrix which diagonalizes the mass matrix ${\bf M}$ in
(\ref{eq:massm1}) is given by
\begin{equation}
U_5 =
\begin{bmatrix}
U_{TBM}  & 0  \\
0  &    I
\end{bmatrix}
\times R_{24}(\theta_{a 1}) \times  R_{35}(\theta_{a 2}),
\label{eq:totmtmix}
\end{equation}
where $R_{24}$ and  $R_{35}$ diagonalize the evolution equations of $(\nu_{a1}, \chi_1)$ 
and $(\nu_{a2}, \chi_2)$ pairs correspondingly. 
Explicitly, we get
\begin{equation}
U_5 =
\begin{bmatrix}
\sqrt{2/3}  &  c_{a 1}/\sqrt{3} & 0                  & s_{a 1}/\sqrt{3}   & 0  \\
-1/\sqrt{6}  &  c_{a 1}/\sqrt{3}  &  c_{a 2}/\sqrt{2} & s_{a1}/\sqrt{3} & s_{a2}/\sqrt{2}   \\
1/\sqrt{6}  &  -c_{a 1}/\sqrt{3}  &   c_{a 2}/\sqrt{2}  & - s_{a 1}/\sqrt{3} & s_{a 2}/\sqrt{2}   \\
0           & -s_{a 1}            &   0                &  c_{a 1}           & 0 \\
0           & 0                 & -s_{a 2}             &  0                & c_{a 2}
\end{bmatrix}\,,
\label{eq:totmtmixex}
\end{equation}
where $c_{a 1, a 2} \equiv \cos \theta_{a 1, a 2}$,  and $s_{a 1, a 2} \equiv \sin \theta_{a 1, a 2}$.

Let us consider bounds arising from active to sterile oscillations of 
solar neutrinos. At low energies, $E < 1 $ MeV, usual matter effect
can be  neglected and the averaged $\nu_e$ survival probability equals
$$
P_{ee} = \sum_j |U_{5ej}|^4 = \frac{5}{9} - \frac{1}{18}
\sin^2 2 \theta_{a 1}.
$$
Thus, for maximal mixing,  $\sin^2 2 \theta_{a 1}=1$,  the oscillations of $\nu_e$ into $\chi$ reduce the
probability by $1/18$ which is within the error bars.
However, the effect at high energies ($^8 B$ neutrinos)
is much stronger:
$P_{ee} \approx 0.5 \sin^2\theta_{12} \approx 0.15$ instead of 0.31, which is excluded.

A way to resolve this problem is to have
$\Delta m^2_{a 1} < 10^{-12}{\,\rm eV}^2$~\cite{deGouvea:2009fp}, 
so that oscillations into sterile components do not develop. In the case of a real field ($\dot{\Phi} = 0$),
this implies, according to Eq. (\ref{eq:splitt}),  that
\begin{equation}
m_{\chi 1} \leq \frac{\Delta m^2_{a1}}{\sqrt{\Delta m^2_{\rm sol}}}
\approx 10^{-10} \, {\rm eV}.
\label{eq:PD_mchi1}
\end{equation}
This bound is shown by the horizontal edge of the hatched region in Fig.\,\ref{fig:Eres_mchi_mphi}.

In the case of maximal asymmetry, the condition of large oscillation length leads to
\begin{equation}
\dot{\Phi} \simeq m_\phi \leq \frac{\Delta m^2_{a1}}{4 E}
\approx 10^{-18} \, {\rm eV}.
\label{eq:PD_Phi}
\end{equation}
The bounds in Eqs.\,(\ref{eq:PD_mchi1}) and (\ref{eq:PD_Phi}) 
are consistent with what we had for refraction in the cold gas.

Another way to satisfy the solar bounds is to arrange for small mixing.
If $\dot{\Phi} \simeq m_\phi$, and $E \dot{\Phi} \gg m_{\chi 1}^2$,
we have, from (\ref{eq:splitt}) and (\ref{eq:mixing}),
\begin{equation}
\Delta m^2_{a 1} \approx 4E\dot{\Phi} = 4 E m_\phi, \, \, \,
\tan 2 \theta_{a 1} =
\frac{m_{\chi 1} \sqrt{\Delta m^2_{\rm sol}}}{E m_\phi},
\label{eq:pseudoDiracMixAng}
\end{equation}
and consequently, from the second equation,
\begin{equation}
\frac{m_{\chi 1}}{m_\phi} < \tan 2 \theta_{a 1}
\frac{E}{\sqrt{\Delta m^2_{\rm sol}}}.
\end{equation}
Requiring  $\tan 2 \theta_{a 1} = 10^{-2}$  for 
$E = 1$ MeV, we obtain
\begin{equation}
\frac{m_{\chi 1}}{m_\phi} <  10^6\,.
\end{equation}
According to Fig. \ref{fig:Eres_mchi_mphi}, this can be satisfied if $m_\phi \gtrsim 10^{-10}$ eV. This bound corresponds to the diagonal border of the hatched region in Fig.\,\ref{fig:Eres_mchi_mphi}.

Notice that for $\dot{\Phi} = 0$, we would get 
\begin{equation}
m_{\chi 1} \geq \frac{2\sqrt{\Delta m^2_{\rm sol}}}{\tan 2 \theta_{a 1}}\,,\,\quad \tan 2\theta_{a 1} 
= 2m_{a 1}/m_{\chi 1}< 10^{-2}\,.
\end{equation}
Therefore, $m_{\chi 1} \geq 1.6$ eV,  and consequently,
 $\Delta m^2_{a 1} \gtrsim 2.6\,$eV$^2$. 
These large values of $m_{\chi 1}$ are, however, in tension with 
bounds arising from big bang nucleosynthesis, as well as from observation of neutrinos 
from SN1987A  and other cosmic sources (discussed in the next sections).

%%%%%%%%%%%%%%%%
\section{Astrophysical and cosmological bounds on parameters}
\label{sec:AstroLab}
%%%%%%%%%%%%%%%%%%%%%%%%%%%%%%%%%%%%%%%%%%%%%%%%%%%%%%%%%%%%
\subsection{Cosmological evolution of refractive mass and structure formation}
\label{sec:cosmomass}
%%%%%%%%%%%%%%%%%%%%%%%%%%%%%%%%%%%%%%%%%%%%%%%%%%%
Decrease of the refractive mass-squared with energy below resonance allows to avoid (at least partially) 
the cosmological bound on the sum of neutrino masses.
Since the DM density redshifts as ordinary non-relativistic matter $\sim (1+z)^3$, 
any neutrino mass sourced from the DM field grows in the early Universe.
The current sum of neutrino masses equals $\sum m_\nu \approx \sqrt{\Delta m^2_{\rm atm}}$ in the case of normal mass hierarchy. 
Therefore, in the period 
of photon decoupling $(z \simeq 1000)$, the neutrino 
mass $m(z)$ becomes as large as $\mathcal{O}(10)\,$eV, thereby making the neutrinos non-relativistic. 
This will be in tension with the 
current Planck limit of  $\sum m_\nu < 0.12\,{\rm eV}$ from observations 
of the CMB as well as baryon acoustic oscillations~\cite{Planck:2018vyg}.
The bound can be evaded for appropriate choice of the resonance energy since for energies below the resonance, 
the refractive neutrino mass decreases with energy and becomes small (see Fig.\,\ref{fig:Delmeff_g_mphi2}), 
if the picture described in sect. \ref{sec:RefMassPart} is correct.

If the refractive mass squared in our solar system is
$\tilde{m}^2$, then the average value of the mass in the Universe 
equals $\tilde{m}^2(0) = \xi\, \tilde{m}^2$, where $\xi^{-1} \simeq 10^5$  is the local overdensity of DM.
Taking into account the energy dependence of the mass, we obtain
at low energies according to (\ref{eq:massvar}),
\begin{equation}
\tilde{m}^2(0) =
\xi \,\tilde{m}^2_{\rm asy}(0)\, y(\epsilon - y)
= \xi\, \tilde{m}^2_{\rm asy}(0)\, \left(\frac{E(0)}{E_R}\right)\,
\left[\epsilon -  \frac{E(0)}{E_R}\right]\,,
\label{eq:present}
\end{equation}
where $E(0)$ is the energy in the present epoch.
Due to redshift,  we have
$\tilde{m}^2_{\rm asy}(z) \propto n_\phi (z) \propto (1 + z)^3$
and $E(z) = E(0) (1 + z)$. Therefore the effective mass squared increased in the past as
\begin{equation}
\tilde{m}^2(z) = \xi \,\tilde{m}^2_{\rm asy}(0)\,  (1 + z)^4\, \left(\frac{E(0)}{E_R}\right)\,
\left[\epsilon -  \frac{E(0)}{E_R} (1 + z) \right]\,.
\label{eq:past}
\end{equation}

If the asymmetry is large  enough, so that the second term
in brackets can be neglected, we obtain
\begin{equation}
\tilde{m}^2(z) = \xi \, \epsilon\,\Delta m_{\rm atm}^2\, (1 + z)^4\,
\left(\frac{E(0)}{E_R}\right)\,,
\label{eq:past1}
\end{equation}
where we used $\tilde{m}^2_{\rm asy}(0) = \Delta m_{\rm atm}^2$.
Eq.\,(\ref{eq:past1}) gives expression
for the resonance energy:
\begin{equation}
E_R = E(0) \,
\xi\, \epsilon\,(1 + z)^4\,\frac{\Delta m_{\rm atm}^2}{\tilde{m}^2(z)} \,.
\label{eq:ER1}
\end{equation}
Using the cosmological bound on sum of neutrino masses
$\tilde{m}^2(z) \leq (\sum m_\nu)^2$,
we find, from (\ref{eq:ER1}), the lower bound
\begin{equation}
E_R \geq E(0)\,\xi\, \epsilon \, (1 + z)^4\,\frac{\Delta m_{\rm atm}^2}{(\sum m_\nu)^2}\,.
\label{eq:ER2}
\end{equation}
For $E(0) = 3\, T_{\rm rel} = 6.9\cdot 10^{-4} $ eV, $z = 1000$ 
and $\sum m_\nu < 0.12\,{\rm eV}$, this equation gives
\begin{equation}
E_R \geq 1.2\, \epsilon\,\, {\rm keV}.
\end{equation}
For zero asymmetry $(\epsilon = 0)$, according to (\ref{eq:past}),
\begin{equation}
\tilde{m}^2(0) = -
\xi\, \tilde{m}^2_{\rm asy}(0)
\left(\frac{E(0)}{E_R}\right)^2\,,
\label{eq:present}
\end{equation}
similar consideration gives
\begin{equation}
\tilde{m}^2(z)= \xi \,\Delta m_{\rm atm}^2\,(1 + z)^5\,\left( \frac{E(0)}{E_R}\right)\, ,
\label{eq:present2}
\end{equation}
and consequently, the bound
\begin{equation}
E_R \geq E(0)\,
\left[\xi \,(1 + z)^5\right]^{1/2}\, \left(\frac{\sqrt{\Delta m_{\rm atm}^2}}{\sum m_\nu}\right)\,.
\label{eq:ER3}
\end{equation}
Numerically, for $z = 1000$, Eq.\,(\ref{eq:ER3}) leads to
$E_R \geq 28$ eV. 
 
Notice that in this consideration we treated
$\tilde{m}^2$ with respect to structure formation as usual VEV mass. 
What matters is the energy density in neutrinos and their group velocity, 
and these characteristics differ in the case of refractive mass.

Indeed, in our case  $E = p + V(p)$,
and the group velocity, following  (\ref{eq:potenwxi1}), equals
\begin{equation}
v_{g} = \frac{dE}{dp} = 1 + \frac{dV}{dp}\,,\qquad
\frac{dV}{dp} \approx 1 -
\frac{\tilde{m}^2_{\rm asy}}{2 E_R^2}
\left[1 - 2\epsilon \frac{p}{E_R} \right]
\label{eq:vgroup}
\end{equation}
in the lowest order in $p/E_R$. This should be compared with
$$
v_g \approx 1  - \frac{m^2}{2 E^2}
$$
in the usual case. Furthermore, the refractive mass squared depends on density perturbation. 
Therefore the cosmological bound should be reconsidered specifically 
for refractive masses taking into account its dependence on energy and density of DM.

%%%%%%%%%
\subsection{Neutrino - Dark matter interactions}
\label{subsec:NDMconstraint}
%%%%%%%%%%%%%%%%%%%%%%%%%%%%%%%%%%%%%%%%%%%%%%%%%%%%%%%%%%%%%%%%
%
Astrophysical neutrinos interact with DM halos as well as DM in the intergalactic 
space along their path to the Earth. This leads 
to energy loss of the neutrinos and, consequently, to a suppression 
of their flux for a given spectra $\propto E^{-\gamma}$ with $\gamma > 0$
~\cite{Barranco:2010xt,Reynoso:2016hjr,Choi:2019ixb,
Ferrer:2022kei,Cline:2022qld,Carpio:2022sml, Cline:2023tkp}. 
This constrains  $\nu-\phi$ interactions, and hence, parameters of our scenario.

\begin{figure}[!t]
       \includegraphics[width=0.45\textwidth]{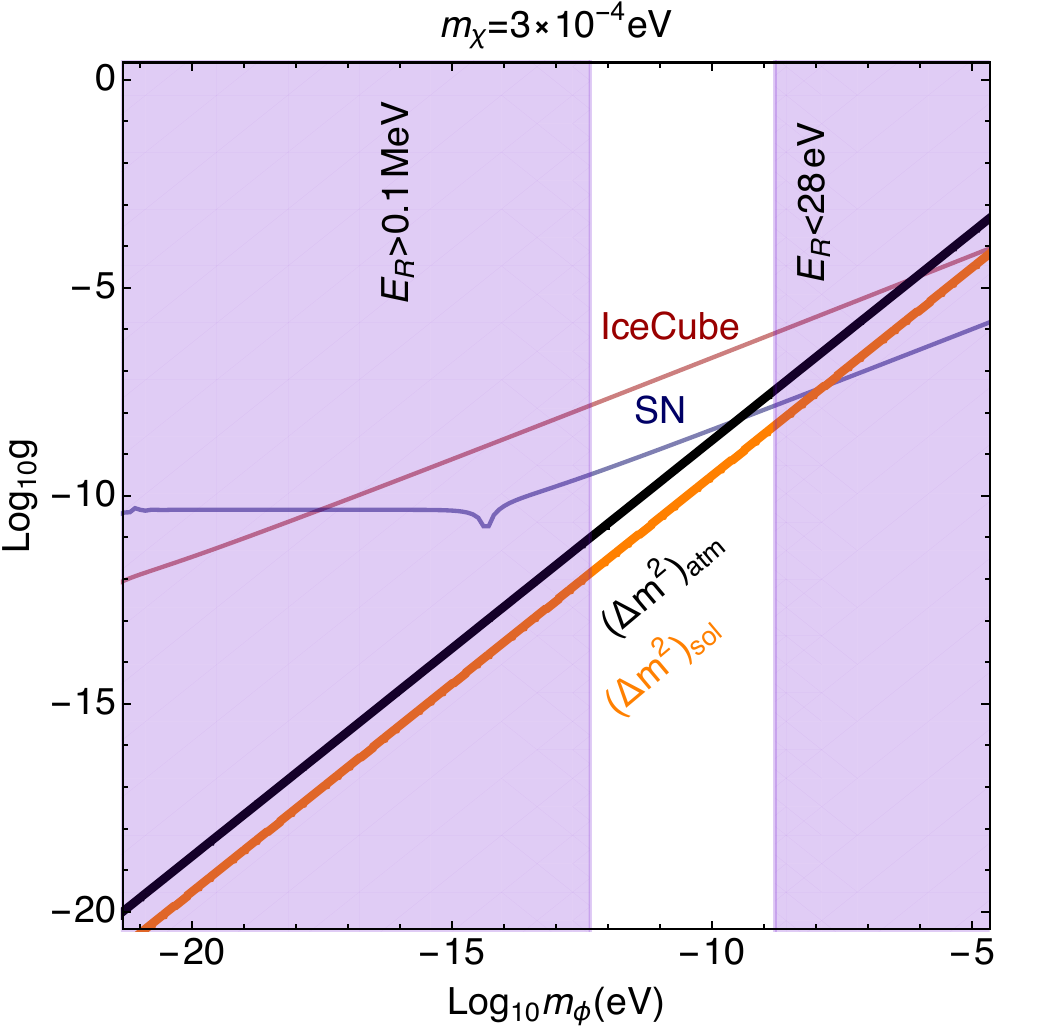}~~\includegraphics[width=0.45\textwidth]{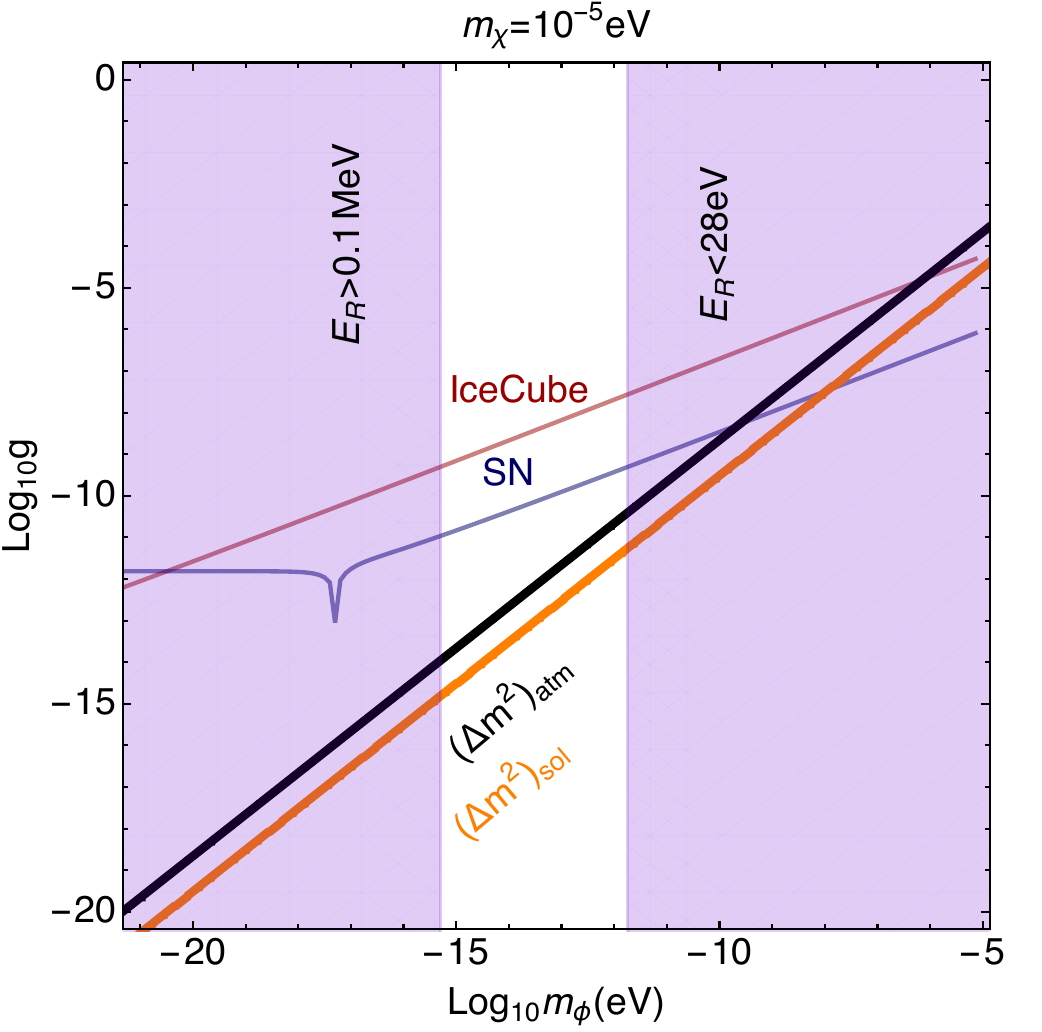}\\
       \vspace{0.1in}
       \includegraphics[width=0.45\textwidth]{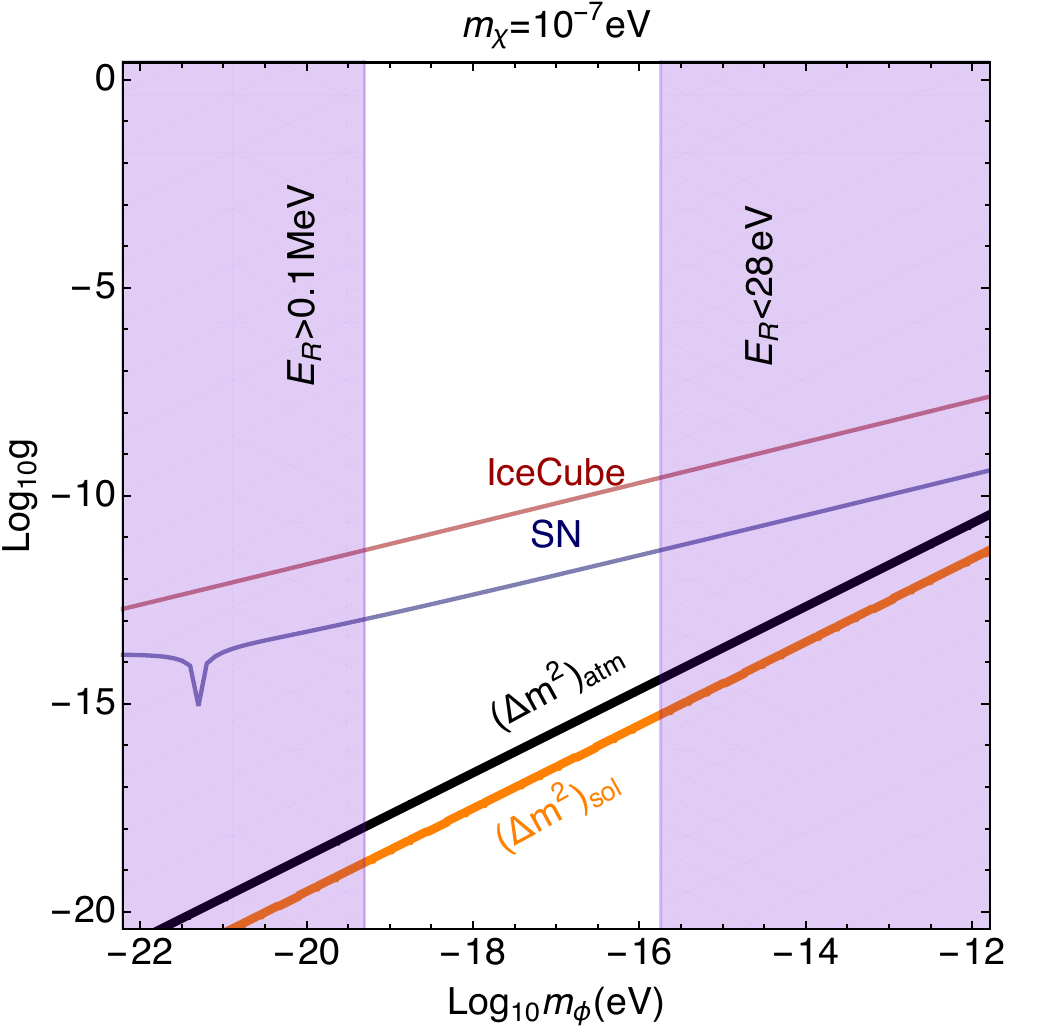}~~
       \includegraphics[width=0.45\textwidth]{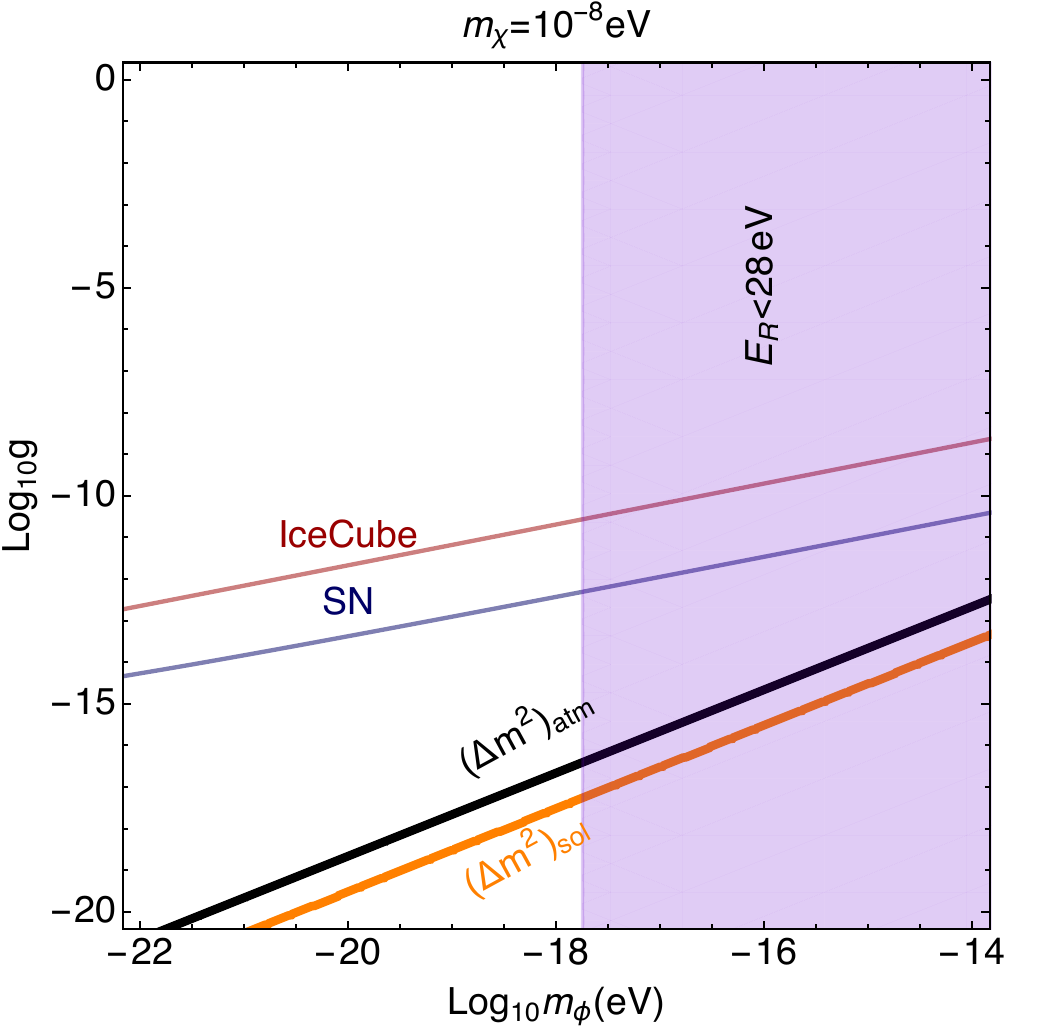}
        \caption{ The bounds and regions required for explanation of oscillation data
by refraction in the $g - m_\phi$ plane for different values of $m_\chi$. 
We take $\rho_\phi=\rho_\odot$. The values of $\Delta m_{\rm atm}^2$  and
$\Delta m_{\rm sol}^2$ are reproduced along the black and orange lines  correspondingly.
Shown are the upper bounds on
$g$ as function of $m_\phi$ from observations of IceCube
high energy neutrinos (red) and neutrinos from SN1987A (blue).
Mauve vertical bands correspond to $E_R > 0.1$ MeV, excluded by oscillation data,  and
$E_R < 28$ eV, excluded by  cosmological bound on sum of neutrino masses.}
\label{fig:bounds}
\end{figure}

The neutrino (inelastic) scattering proceeds due to $\chi$ exchange 
on $\phi^*$ ($s-$channel) and $\phi$ ($u-$channel) and is described by the same diagrams as 
refraction (see Fig.\,\ref{fig:potential}). However, in contrast to refraction, these contributions sum up incoherently. 
Below and above the resonance energy, the s-channel cross-section can be written as  
\begin{equation}
  \sigma_{\nu\bar{\phi}}  =
    \begin{cases}
      \frac{g^4 m_\phi E}{16 \pi m_\chi^4 }\,,  & m_\phi \ll E  \ll  E_R \,,  \\
      \frac{g^4 }{64 \pi m_\phi E }\,, &  E \gg E_R\,.
    \end{cases}       
    \,
    \label{eq:sigmaslim}
\end{equation}
For the $u-$channel, the cross section is given by
\begin{equation}
\sigma_{\nu\phi} =
 \begin{cases}
 \frac{g^4 m_\phi E }{8 \pi m_\chi^4 }\,, &  m_\phi\ll E \ll  E_R\,, \\
\frac{g^4}{32 \pi m_\phi E }\left[-1+{\rm log}\left(\frac{2 E m_\phi}{m_\chi^2}\right)\right]\,, & E \gg E_R \, . 
 \end{cases}       
    \,
\label{eq:sigmaulim}
\end{equation}
The optical depth of neutrinos in $\phi-$ background equals
\begin{equation}
\label{eq:OptDep}
\tau \equiv \int (n_\phi \sigma_{\nu \phi} + \bar{n}_{\phi} \sigma_{\nu\bar{\phi}}  )\, dl_{\rm los}\, 
=\frac{1}{m_\phi} \int_0^L \left( \rho_\phi \sigma_{\nu \phi} + 
\bar{\rho}_\phi\sigma_{\nu{\bar{\phi}} } \right) \, dl_{\rm los}\, , 
\end{equation}
The integration proceeds along the line of sight $l_{\rm los}$ from the source to the Earth.  
Therefore, the experimental bound on suppression 
of the neutrino flux from the source at distance  $L$ is transformed into a bound on the ratio of the total 
cross-section and $m_\phi$. This, in turn, gives an upper bound  on $g$ as function of $m_\phi$ for different values of $m_\chi$.   
To obtain such a dependence for a given experiment, 
which detects neutrinos of the energy $\sim E$, we define the resonance value of $m_\phi$: 
\begin{equation}
m_\phi^R \equiv \frac{m_\chi^2}{2E}\,.
\end{equation}
Notice that with increase of energy $E$, the mass $m_\phi^R$ shifts to smaller values. Then Eqs. (\ref{eq:sigmaslim}) and (\ref{eq:sigmaulim}) 
give the upper bounds on $g$ as a function of the parameters 
$m_\phi$, $m_\chi$ and energy as follows.
For a fixed $\rho$, in the range $m_\phi \ll m_\phi^R$, which corresponds to 
$E \ll E_R$, we find that  $g \propto m_\chi/E^{1/4}$, while for $m_\phi \gg m_\phi^R$ ($E \gg E_R$), we get $g \propto m_\phi^{1/2} E^{1/4}$.
These dependences can be seen in the IceCube and SN constraints in Fig.\,\ref{fig:bounds}, which will be discussed below.

%%%%%%%%%%%%%%%%%%%%%%%%%%%%%%%%%%%%%%%%%%%%%%%%
%%%%%%%%%%%%%%%%%%%%%%%%%%%%%%%%%%%%%%%%%%%%%

%%%%%%%%%%%%%%%%%%%%%%%%%%%%%%%%%%%%%%%%%%%%%%%%%%%%%%%%%%%%%%%%%%%%%%%%%%%%%

In our computations, we use the DM density distribution in the Milky Way described by 
a Navarro-Frenk-White (NFW) profile~\cite{Navarro:1996gj}.
The cosmological DM density in the Universe redshifts like as 
$\rho_{\rm EG}(z)=\xi \rho_\odot  (1+z)^3$.
The total DM density is given as a sum of these 
two terms, that is, $\rho_\phi + \bar{\rho}_\phi =
\rho_{\rm gal} +  \rho_{\rm EG} $. Assuming that the spatial distributions 
of $\phi$  and  $\bar{\phi}$ are the same $(\rho_\phi \sim  \bar{\rho}_\phi)$, 
we can write expression for the optical depth as 
\begin{equation}
\label{eq:OptDep2}
\tau  = \frac{1}{2 m_\phi} 
\left[(1 + \epsilon)\sigma_{\nu \phi} +  (1 - \epsilon) \sigma_{\nu{\bar{\phi}}} \right]
\int \left( \rho_{\rm gal} + \rho_{\rm EG}  \right) \, dl_{\rm los}\, .  
\end{equation}

For a symmetric medium ($\epsilon = 0$),
the line-of-sight integral over the galactic component $\rho_{\rm gal} $ yields
\begin{equation}
\label{eq:Gal}
\int  \rho_{\rm gal}\,  dl_{\rm los}= 2.6\times10^{22}\, ~{\rm GeV}/{\rm cm}^2\,. 
\end{equation}
The extra-galactic contribution to $\tau$  is given by integration  over the redshift as
\begin{equation}
\int  \rho_{\rm EG}\, dl_{\rm los}= \int_0^{z_0} \frac{\rho_{\rm EG} }{H_0 (1+z)
\sqrt{\Omega_\Lambda + \Omega_m(1+z)^3}}\, dz\,,
\end{equation}
where $H_0=67.4$ km/s/Mpc, $\Omega_\Lambda=0.685$ and $\Omega_m=0.315$~\cite{Planck:2018vyg}.

\subsection{Bounds from SN1987A and IceCube neutrinos}
%%%%%%%%%%%%%%%%%%%%%%%%%%%%%%%%%%%%%%%%%%%%%%%%%
Let us apply the formulae of subsection\,\ref{subsec:NDMconstraint} to existing observations of astrophysical neutrinos. 

1. SN1987A. 
Requiring that the suppression of the $\mathcal{O}(10)\,{\rm MeV}$ neutrino 
flux from SN1987A~\cite{Hirata:1987hu,Hirata:1988ad,Bionta:1987qt,Alekseev:1988gp} 
at a distance of around $50\,$ kpc  from the Earth is no more than a factor 0.1 gives  $\tau \leq 2.3$.
Using only $\rho_{\rm gal} $ in Eq. (\ref{eq:OptDep2}), we obtain
\begin{equation}
\frac{\sigma_{\nu \phi} + \sigma_{\nu{\bar{\phi}} } }{m_\phi}  \lesssim   9\times10^{-23} \,{\rm cm}^2/{\rm GeV}\,.
\label{eq:SN_Constraint}
\end{equation}
%%%

2. IceCube. 
IceCube has reported the observation of a $290\,$TeV cosmic neutrino 
in association with the $\gamma$-ray blazar TXS 0506+056 at a redshift $z_0=0.3365\pm 0.0010$, 
corresponding to a distance of $~1421\,$Mpc~\cite{IceCube:2018dnn}.  
Now, the requirement  $\tau \leq 2.3$ gives
\begin{equation}
\frac{\sigma_{\nu \phi} + \sigma_{\nu{\bar{\phi}} } }{m_\phi}  \lesssim   7\times 10^{-23} ~ \,{\rm cm}^2/{\rm GeV}\,.
\label{eq:NDM_Constraint}
\end{equation}

Using the expressions for $\sigma_{\nu \phi}$ and $ \sigma_{\nu{\bar{\phi}}}$, 
we convert the bounds (\ref{eq:SN_Constraint}) and (\ref{eq:NDM_Constraint}) 
into upper bounds on $g$ as a function of $m_\phi$ (see Fig.\,\ref{fig:bounds}).
For simplicity, we assume that $m_{\chi 1} =m_{\chi 2}=m_\chi$. The bounds  agree with those obtained in \cite{Choi:2019ixb}.
Along the black and orange lines, the observed solar and atmospheric neutrino 
mass-squared difference can be obtained using refractive masses only.

The vertical shaded regions in mauve show the constraints arising 
from the condition $28\,{\rm eV}\leq E_R \leq 0.1\,{\rm MeV}$, as discussed in Sec.\ref{sec:refmass}. 
According to Fig.\,\ref{fig:bounds}, values of the parameters which satisfy the existing bounds 
and allow to reproduce oscillation data are in the following ranges:
\begin{equation}
m_\chi  =  \left(3\cdot 10^{-9} \div 10^{-4}\right) ~{\rm eV},~~~ m_\phi = \left(10^{-19} \div 10^{-10}\right) ~{\rm eV},
~~~ g = 10^{-18} \div 10^{-7}\,.
\end{equation}

%%%%%%%%
\subsection{Cosmological constraints on $\nu - \chi$ mixing}
%%%%%%%%      
\label{sec:Cosmo} 
%%%%%%%%%%%%%%%%%%%%%%%%%%%%%%%%%%%%%%%                   
%%%%%%%%%%
%%%%%%%%%%%%%%%%%%%%%%%%%%%%%%%%%%%%%%%%%%%%%%%%%%%%%%%%%%%%%%%%
                                        
The relic density of $\phi$ can be set by the misalignment mechanism~\cite{Preskill:1982cy,Abbott:1982af,Dine:1982ah}. 
According to this mechanism, $\phi$ is initially displaced from the minimum of the potential. 
The field remains constant due to Hubble friction. 
It starts oscillating at temperatures below $T_H$, determined by the equality  
$m_\phi=3H(T_H)$, where  $H$ is the Hubble parameter. 
For $T<T_H$, the DM redshifts as non-relativistic matter
so that its energy density varies as $T^3$ ~\cite{Dev:2022bae}: 
\begin{equation}
\rho_\phi(T)=\rho_\phi (T_0) \left(\frac{g_*(T_0)}{g_*(T)}\right)\left(\frac{{\rm min}(T,T_H)}{T_0}\right)^3\, . 
\end{equation}
Here $T_0=2.72\,$K is the present temperature of the CMB.
Consequently, the DM density increases with temperature 
upto $T$ at which $m_\phi=3H$, and then remains constant.

In the presence of the classical field $\phi_c$, the mediator $\chi$ 
mixes with $\nu$ and therefore can be produced from $\nu - \chi$ oscillations in the early Universe through the Dodelson-Widrow mechanism~\cite{Dodelson:1993je}. The distribution of $\chi$ can be computed through the following semi-classical Boltzmann equation,
%%%%%%%%%%                                                                                                   
\begin{eqnarray}
\label{eq:masterequation}
\frac{ \mathrm{d} f_{\chi i} } {\mathrm{d} T} & =
- & \frac{\Gamma_\nu}{4H T} \frac{ \left(\Delta m^2 _{a i}   \sin\,2\theta_{a i}\right)^2  }{  \left(\Delta m^2 _{a i}   \sin\,2\theta_{a i}\right)^2
  + E^2\Gamma_\nu^2+ \left( \Delta m^2_{a i} \cos 2 \theta_{a i}  -2 E V \right)^2}\,\, f_{\nu_a} \,.
\end{eqnarray}
Here $\Delta m^2_{a i}$ and $\theta_{a i}$ are oscillation parameters
in the $(\nu_a',\chi_i)$ system as defined in (\ref{eq:splitt}), 
(\ref{eq:mixing}). 
In Eq. (\ref{eq:mixing}), the thermal scattering rates of neutrinos $\Gamma_\nu$
and the effective forward scattering matter potential in a symmetric medium are given by~\cite{Abazajian:2001nj}
\begin{eqnarray}
\Gamma_\nu &=& \frac{7 \pi }{24} G_F^2 E T^4\,,\\
        V    &=&-\frac{14\sqrt{2}\pi^2}{45} G_F E T^4\left(\frac{1}{M_Z^2} + \frac{2}{M_W^2} \right)\,.
\end{eqnarray}

Eq. (\ref{eq:masterequation}) can be solved numerically to obtain the contribution of $\chi$
to extra radiation around the time of big-bang nucleosynthesis (BBN):
\begin{equation}
  \Delta N_{\rm eff}=\frac{\rho_\chi }{\rho_{\nu}}\,,
\end{equation}
where $\rho_{\chi,\nu}$ is the energy density associated with $\chi$ and $\nu$ respectively.  The current state-of-the-art limit
is $\Delta N^{\rm BBN}_{\rm eff} <  0.5$ \cite{Gariazzo:2021iiu}.

The choices of $g_{e2}=0$ in the TBM form imply 
that the $\nu_e-\chi_2$ mixing is zero, $\sin 2\theta_{e2}=0$, therefore $\chi_2$ 
cannot be produced by $\nu_e$ through oscillations; however, 
oscillations $\nu_{\mu,\tau}\to \chi_2$ can happen. Similarly, $\chi_1$ can be produced 
by all three neutrino flavours in the early Universe. 
This can be suppressed by requiring that $\nu-\chi$ oscillations do not develop 
before $T\sim 1\,{\rm MeV}$, thereby avoiding the BBN bounds. 
This can be satisfied if $\Delta m^2_{ai}/(2T) < H$ at around $T=1\,{\rm MeV}$. 
This gives $\Delta m^2_{ai}<10^{-9}\,{\rm eV}^2$. For a real $\phi$, 
according to (\ref{eq:PD_mchi1}), this leads to $m_\chi<10^{-9}\,{\rm eV}$, assuming that the refractive mass stops growing at $z=1000$ and is constant till the BBN epoch. This bound is weaker than that from solar neutrinos.
Another way of evading the BBN bound is to consider small mixing angles $\theta_{ai}$, as given in 
(\ref{eq:pseudoDiracMixAng}). This would also prevent $\chi$ from being populated in early Universe from $\nu-\chi$ mixing.

Other constraints can exist for the classical field component of ultralight dark matter. 
The wavelike nature of such a dark matter candidate can lead to wave interference, 
causing fluctuations which heat up stellar objects. Observations of ultrafaint dwarf (UFD) 
galaxies have been used to rule out masses of fuzzy dark matter $m_\phi<10^{-19}\,{\rm eV}$~\cite{Dalal:2022rmp}. 
However, this constraint can be evaded by allowing for $\phi$ to produce only a fraction 
of the DM relic density. Hence we do not discuss these bounds further.

Combining both astrophysical and cosmological bounds considered in this section (see also Fig.\,\ref{fig:Eres_mchi_mphi} and Fig.\,\ref{fig:bounds}), we 
obtain viable ranges of parameters,
\begin{equation}
m_\chi  \simeq  (1 \div   3)\cdot 10^{-4}     ~{\rm eV},~~~ m_\phi \simeq  10^{-10} \div 10^{-9} ~{\rm eV},
~~~ g \simeq 10^{-10} \div 10^{-9}\,.
\end{equation}
They are in agreement with the numbers found in~\cite{Choi:2020ydp}. 
Another possible region is given by
\begin{equation}
m_\chi  \simeq  10^{-10}     ~{\rm eV},~~~ m_\phi \simeq  10^{-21} ~{\rm eV},
~~~ g \simeq 10^{-20}\,.
\end{equation}

%%%%%%%%%%%%%%%%%%%%%%%%%%%%%%%%%%%%%%%%%%%%%%%%%%%%%%
\section{Discussion and Conclusion}
\label{sec:Concl}
%%%%%%%%%%%%%%%%%%%%%%%%%%%%%%%%%%%%%%%%%%%%%%%%%%%

We explored in detail a possibility that neutrino oscillation results
can be explained by interactions of massless neutrinos with ultralight
scalar dark matter $\phi$.  The scheme includes light fermionic mediator $\chi$  
with mass $m_\chi \gg m_\phi$ and Yukawa coupling
$g \bar{\chi} \nu \phi^* + h.c. $.  Explanation of neutrino oscillations require the existence of at least two mediators.

We introduced the refractive neutrino mass squared, $\tilde{m}^2$, generated dynamically 
via refraction of neutrinos in the $\phi-$background.
Properties of the refractive masses and their
phenomenological consequences are studied.
We obtained constraints on parameters of the scheme $(g, m_\phi, m_\chi)$ from  neutrino oscillations, 
as well as astrophysical and cosmological observations.

Properties of the refractive mass depend
on the state of the scalar background, in particular, on whether it appears as
a cold bath of scalar particles  or as
a classical scalar field representing a coherent state of scalar bosons.

In the case of a cold bath of bosons, the refraction has a resonance dependence on energy related to $\chi$ production.
Above the resonance, $E>E_R=m_\chi^2/(2 m_\phi)$, the refractive mass has the same
properties as the usual vacuum mass: it does not depend on energy, being equal for
neutrinos and antineutrinos. $E_R$ should be much smaller than the lowest energy of detected neutrinos for which no dependence
of masses and mixing on energy is found. This gives the upper bound $E_R \leq 0.1\,{\rm MeV}$.
Below resonance $(E<E_R)$, the refractive mass decreases with energy.
For C-asymmetric background, it decreases as $\epsilon/E$, where $\epsilon$ quantifies the asymmetry,
while in a C-symmetric medium (real field), the decrease is stronger: $1/E^2$.
This can lead to very small (unobservable) effective neutrino mass in beta-decay experiments like KATRIN.

The key difference of the refractive mass from the usual VEV-induced mass is that it is proportional to the number
density of particles of the background and therefore its value increased in the past due to redshift.
Decrease of $\tilde{m}^2$ with energy below the resonance allows to satisfy the
cosmological bound on the sum of neutrino masses obtained
from structure formation. The bigger the resonance energy, the stronger is the decrease of mass.
For C-symmetric background we obtain from cosmology the lower bound
$E_R  > 28\,{\rm eV}$. Thus, $E_R \in [30 - 10^5]$ eV.
For completely asymmetric background $(\epsilon=1)$, the lower bound is much stronger:  $E_R > 1.2\,{\rm keV}$.

For a fixed value of the mediator mass $m_\chi$, the allowed interval for the resonance energy 
is transformed onto the interval for the scalar mass. The bounds on the coupling $g$
as function scalar mass $m_\phi$ follow from
scattering of astrophysical neutrinos  (high energy cosmic neutrinos and neutrinos from SN1987A) on the background scalars.
This scattering leads to energy loss of the neutrinos,
and suppresses their flux. Combining both astrophysical and cosmological bounds constrain the viable ranges of parameters as
$
m_\chi  \simeq  (1 \div   3)\cdot 10^{-4}     ~{\rm eV},~~~ m_\phi \simeq  10^{-10} \div 10^{-9} ~{\rm eV},
~~~ g \simeq 10^{-10} \div 10^{-9}\,.
$
Another possible region is given by
$
m_\chi  \simeq  10^{-10}     ~{\rm eV},~~~ m_\phi \simeq  10^{-21} ~{\rm eV},
~~~ g \simeq 10^{-20}\,.
$
For these values of parameters, the refractive mass can reproduce the value $\Delta m^2_{\rm atm, sol}$ in the 
lowest order of perturbation theory,

Very small mass of $\phi$ means that these particles have a much larger de-Broglie wavelength
than the radius of interaction, which in turn, is much larger than the distance between the scatterers.
This indicates that multiple neutrino interactions with $\phi$ may become important.
We find that for parameters required to reproduce the observed neutrino masses,
the perturbation theory related to multiple $\phi$ interactions 
is broken at low energies. The perturbation parameter $\zeta$ increases with decrease in energy as $\zeta \sim 1/E$,
and $\zeta (E_{\rm p}) \simeq 1$ is achieved at energies above resonance: $E_{\rm p} \gg E_R$.
For a C-symmetric medium, $E_{\rm p}$ can be smaller than 0.1 MeV so that 
results obtained for $\tilde{m}^2$ in the lowest approximation remain valid. 
Perturbativity can break down for energies close to the resonance energy. 
Below resonance, $\zeta$  is independent of energy,
so we expect the refractive neutrino mass to decrease with energy as before,
thus satisfying the cosmological bound. Qualitatively, in the non-perturbative situation, 
the dependence of $\tilde{m}^2$ on $E$ can be similar to the perturbative one in the lowest order. 
Also due to the $\bar{\chi} \nu \phi $ coupling,  the off-diagonal $\chi^c - \nu$ potentials are generated, which lead to $\chi^c - \nu$ mixing and oscillations.

Notice that at the condition $d \ll r_\chi$, the standard computations 
of the local potentials $V_{\alpha, \beta}$ with integration over the infinite spatial 
coordinates may become problematic. 

In the case of a background described by classical scalar field (coherent state of scalar bosons), 
the coupling $g \bar{\chi} \nu \phi^*$ generates  off-diagonal mass terms for
$\nu$ and $\chi$, that is, the off-diagonal elements of the mass matrix $M$.
Then the mass matrix squared,  $M M^\dagger$, has the $3 \times 3$ active neutrino block proportional
to that obtained for refractive mass squared at high energies.
This mass has no explicit energy dependence and therefore differs from refractive mass at low energies. However,  the $\phi$ field, and consequently, the neutrino
mass squared have time dependence which effectively
reproduces the resonance features.
The resonance in the refractive mass corresponds to resonance in $\nu - \chi$
oscillations.

We find that at high energies, the Hamiltonian of evolution of whole the system of $(\nu, \chi^c)$ is essentially 
the same for coherent classical field and a cold gas of scalars. 

The $\nu - \chi$ mixing and oscillations are relevant for neutrinos from astrophysical sources. 
The strongest constraints arise from solar neutrinos. This can be evaded in two ways. 
For $m_\chi<\sqrt{\Delta m^2_{\rm sol}}$,
neutrinos are pseudo-Dirac and the mixing between $\nu$ and $\chi$ is maximal. 
In this case, the solar neutrino constraint can be evaded by not allowing $\nu-\chi$  oscillations to develop over the solar baseline.
For a real $\phi$, this imposes the bound
 $m_\chi < 10^{-10}\,{\rm eV}$.  Another way of avoiding these bounds is to arrange for small $\nu-\chi$ mixing.
The mixing is given by $m_\chi \sqrt{\Delta m^2_{\rm sol}}/( E m_\phi)$ and can be
suppressed for $m_\chi < 10^{6}\, m_\phi$ at $E=1\,{\rm MeV}$. 
This condition is consistent with the other constraints for $m_\phi\gtrsim 10^{-10}\,{\rm eV}$.

With the identification $\phi=\sqrt{n_\phi/m_\phi}$, both the states of the background 
can lead to the same effective neutrino masses at observable energies. 
In the case of a real $\phi$ or in the presence of a coherent state of $\phi$ and $\phi^*$, the 
neutrino refractive masses can show time-variations. 

The neutrino oscillation phenomena are directly related to refraction on the dark matter background.
The neutrino mass can change periodically with time and can be different in different spatial points.
It may show energy dependence at low energies. All these features should be subject of
experimental searches.

The masses of the new fermions $\chi$ need to be introduced by some high-scale physics.
It may be related to some Planck scale physics, so that $m_\chi \sim v_{EW}^2/M_{Pl}$.
Alternatively it can be generated by some refraction effect in the dark sector.

In conclusion, the nature of neutrino mass may be related to the nature of dark matter and the cosmological evolution of the Universe.

\section*{Acknowledgement}

%%%%%%%%%%%%%%%%%%%%%%%%%%%%%%%%%%%%%%%%%%%%%%%%%%%%%%%%%%%%%%%%%%%%%%%%
The authors are thankful to E. Kh. Akhmedov,  Sacha Davidson, G. Huang,  J. Herms, J. Jaeckel,  G. Raffelt, A. Trautner, and G. Villadoro for useful discussions.
A.Yu. S.  appreciates very much numerous discussions with Eung Jin Chun, Ki-Young Choi and Jongkuk Kim during his stay at KIAS.

%%%%%%%%%%%%%%%%%%%%%%%%%%%%%%%%%%%%%%%%%%%%%%%%%%%%%%%%%%%%%%%%%%%%%%%%%%%%%%%%%%%%%%%%%%%%%%%%%

%\newpage

%\newpage
\appendix
\section{General formalism for a coherent field}
\label{sec:Appendix}
%%%
%%%
In this appendix, we derive the general formalism for a coherent field. We  consider a complex scalar field operator
\begin{equation}
\hat{\phi}(x) = \int \frac{d^3 q}{ (2\pi)^3} \frac{1}{\sqrt{2 E_q}}
\left(\hat{a}_q e^{- i q x}  + \hat{b}^{\dagger}_q e^{i q x} \right).
\label{eq:scalfildcompApp}
\end{equation}
and its hermitian conjugate $\hat{\phi}^\dagger (x)$. The coherent states  can be constructed  in the second quantization formalism using creation and annihilation
operators as
\begin{equation}
|\phi_{\rm coh} \rangle = \frac{1}{A} \exp \left\{
\int \frac{d^3 {k}}{ (2\pi)^3}
\left(f_a({ k}) \hat{a}^\dagger_k +  f_b({k}) \hat{b}_k \right)
\right\} |0 \rangle ,
\label{eq:chfieldApp}
\end{equation}
\begin{equation}
|\bar{\phi}_{\rm coh} \rangle = \frac{1}{A} \exp \left\{
\int \frac{d^3 { k}}{ (2\pi)^3} 
\left(\bar{f}_a({ k}) \hat{a}_k +  \bar{f}_b ({ k}) \hat{b}_k^\dagger \right)
\right\} |0 \rangle .
\label{eq:chfieldantApp}
\end{equation}
Here  $A$ is the normalization factor such that   $\langle \phi_{\rm coh} |\phi_{\rm coh} \rangle = 1$,
and
$f_i({ k})$ are the spectra of scalars (weight function for the mode ${ k}$).
A coherent field appears as expectation value of the field operator in the coherent state of particles:
\begin{equation}
\phi_c (x) = \langle \phi_{\rm coh} |\hat{\phi}(x) |\phi_{\rm coh} \rangle.
\label{eq:expvalApp}
\end{equation}
Let us consider the most general case
\begin{equation}
|{\phi}_{\rm coh}^{\rm tot} \rangle = \cos\alpha |{\phi}_{\rm coh} \rangle + \sin \alpha |\bar{\phi}_{\rm coh} \rangle .
\label{eq:totalstApp}
\end{equation}
Inserting (\ref{eq:totalstApp}),  (\ref{eq:scalfildcompApp}) and (\ref{eq:chfieldApp})
into (\ref{eq:expvalApp}), we obtain
\begin{equation}
\phi_c^{\rm tot}(x) = \langle{\phi}_{\rm coh}^{\rm tot}|\hat{\phi}(x) |{\phi}_{\rm coh}^{\rm tot} \rangle
= \int \frac{d^3k}{ (2\pi)^3} \frac{1}{\sqrt{2 E_k}}
\left[f_a^{\rm tot}({ k}) e^{-i k x} + f_b^{\rm tot}({ k}) e^{i k x} \right],
\label{eq:phiccomApp}
\end{equation}
where 
\begin{equation}
f_a^{\rm tot}  = \cos^2 \alpha f_a + \sin^2 \alpha \bar{f}_a^\dagger, ~~~~~
f_b^{\rm tot}  = \cos^2 \alpha f_b + \sin^2 \alpha \bar{f}_b^\dagger.
\label{eq:ffaffbApp}
\end{equation}
The field $\phi_c^{\rm tot}(x)$ can be parametrized as
\begin{equation}
\phi_c^{\rm tot}(x) =  \phi_a e^{-i\theta_a} +  \phi_b e^{i\theta_b}.
\label{eq:parampApp}
\end{equation}
Here
\begin{eqnarray}
\phi_a e^{-i\theta_a} & = & \int \frac{d^3k}{ (2\pi)^3} \frac{1}{\sqrt{2 E_k}}
f_a^{\rm tot}({ k}) e^{-i k x},\\
\phi_b e^{i\theta_b} & = & \int \frac{d^3k}{ (2\pi)^3} \frac{1}{\sqrt{2 E_k}}
f_b^{\rm tot}({ k}) e^{i k x}.
\end{eqnarray}
The field (\ref{eq:parampApp}) can be written as
\begin{equation}
\phi_c^{\rm tot}(x) = F(x) e^{i\Phi} \, ,
\end{equation} 
such that 
\begin{eqnarray}
F(x)^2&=& \phi_a^2 +\phi_b^2 + 2\phi_a \phi_b  \cos(\theta_a  + \theta_b)\,,\\
\tan\Phi &=& \frac{-\phi_a \sin\theta_a + \phi_b \sin\theta_b}{\phi_a \sin\theta_a + \phi_b \sin\theta_b}\,.
\end{eqnarray}
For the Hermitian conjugate we have similarly
\begin{equation}
\phi_c^{\rm tot~\dagger}(x) = \langle{\phi}_{\rm coh}^{\rm tot}|\hat{\phi}^\dagger(x)|{\phi}_{\rm coh}^{\rm tot} \rangle =
\phi_a^\dagger e^{i\theta_a} +  \phi_b^\dagger e^{-i\theta_b} \equiv \bar{F}(x) e^{-i \bar{\Phi}}\,.
\label{eq:parampdagApp}
\end{equation}
For a real field, we have $\phi_a=\phi_b=\phi_0$ and $\theta_a=\theta_b=\theta$. This reduces to the usual results for a real field and consequently
\begin{eqnarray}
F(x)&=& 2\phi_0 \cos(\theta)\,,\\
\tan\Phi &=& 0\,.
\end{eqnarray}
On the other hand, for a complex field, in the highly non-relativistic approximation $k\simeq (m_\phi,0)$, we have
\begin{eqnarray}
F(x)^2&=& \phi_a^2 +\phi_b^2 + 2\phi_a \phi_b  \cos(2 m_\phi t)\,,\\
\tan\Phi &=& \frac{-\phi_a  + \phi_b}{\phi_a  + \phi_b} \tan m_\phi t\,.
\end{eqnarray}
The time variation of $\Phi$ that appears in the Hamiltonian equals
\begin{equation}
\dot{\Phi} = \frac{\eta\, m_\phi}{1-(1-\eta^2)\sin^2 m_\phi t}\,,
\end{equation}
where 
$$
\eta= \frac{-\phi_a  + \phi_b}{\phi_a  + \phi_b}\,.
$$
Here $\dot{\Phi} \in [0, m_\phi]$.

\bibliographystyle{JHEP}
\bibliography{biblio}

\end{document}